\documentclass[11pt,twoside]{article}
\usepackage{graphicx}
\begin{document}
%%%%%%%%%%%%%%%%%%%%%%%%%%%%%%%%%%%%%%%%%%%%%%%%%%%%%%%%%%%%%%%%%%%%%%%%%%%%%%%%%%%

\title{Linear algebra and differential geometry on abstract Hilbert space}
\author{Alexey A. Kryukov
\footnote{Department of Mathematics, University of Wisconsin Colleges
\newline
 E-mail: alexey.kryukov@uwc.edu, aakrioukov@facstaff.wisc.edu}
}
\maketitle

\pagestyle{myheadings} 
\thispagestyle{plain}         
\markboth{Alexey A. Kryukov}{Linear algebra and differential geometry on abstract Hilbert space} 
\setcounter{page}{1}

\begin{abstract}
Isomorphisms of separable Hilbert spaces are analogous to isomorphisms of $n$-dimensional vector spaces. However, while $n$-dimensional spaces in applications are always realized as the Euclidean space $R^n$, Hilbert spaces admit various useful realizations as spaces of functions. In the paper this simple observation is used to construct a fruitful formalism of local coordinates on Hilbert manifolds. Images of charts on manifolds in the formalism are allowed to belong to arbitrary Hilbert spaces of functions including spaces of generalized functions. Tensor equations then describe families of functional equations on various spaces of functions. The formalism itself and its applications in linear algebra, differential equations and differential geometry are analyzed.
\end{abstract}

\bigskip

%%%%%%%%%%%%%%%%%%%%%%%%%%%%%%%%%%%%%%%%%%%%%%%%%%%%%%%%%%%%%%%%%
%\newpage
\section{Introduction}
\setcounter{equation}{0}

Various integral transforms are known to be extremely useful in analysis and applications. One example is the Fourier transform in analysis and applied problems; another example is the Segal-Bargmann transform \cite{Barg} in quantum theory.
An important common property of integral transforms is that they relate various spaces of functions and various operators on these spaces and allow one to ``transplant'' a problem from one space to another one. Because of that, the problem at hand may become easier to solve. 
A somewhat similar situation arises when working with tensor equations in the finite dimensional setting. Namely, by an appropriate choice of coordinates one can significantly simplify a given tensor equation. Although the analogy is obvious, an infinite-dimensional setting offers a significantly larger variety of situations. In particular, by using the Segal-Bargmann transform, one can relate problems on spaces of ordinary or even generalized functions to problems on spaces of holomorphic functions.

In the paper we attempt to build a systematic approach to functional transformations on Hilbert spaces based on the above mentioned analogy between integral transforms and changes of coordinates on an $n$-dimensional manifold. Initial results in this direction were announced in \cite{Kryukov1} and  \cite{Kryukov}. Various application of the formalism to quantum theory especially to the problem of emergence of the classical space-time were considered in \cite{Kryukov}, \cite{Kryukov3} and \cite{Kryukov4}. The goal here is to approach the subject in a more formal mathematical way and to justify the previously obtained results. The reader is referred to \cite{Lok} for an introduction to Hilbert spaces and applications.
 
We begin by describing a method of building various Hilbert spaces of functions including spaces of smooth and generalized functions. Then a simple formalism based on isomorphisms of these spaces is constructed. The formalism allows one to ``move'' in a systematic way between Hilbert spaces of functions and at the same time to formulate problems in a way independent of any particular functional realization. The formalism is then applied to building linear algebra on Hilbert spaces which deals with the ordinary and the generalized functions on an equal footing. After discussing various special transformations preserving properties of differential operators we concentrate on differential geometry of Hilbert spaces of functions. Namely, using the developed formalism we find a natural isometric embedding of finite dimensional Riemannian manifolds into Hilbert spaces of functions. Finally, the formalism is applied to construct a Riemannian metric on the unit sphere in a Hilbert space in such a way that solutions of Schr{\"o}dinger equation are geodesics on the sphere.

%\newpage

\section{Hilbert spaces of $C^{\infty}$-functions and their duals}
\setcounter{equation}{0}

Let us discuss first a general method of constructing various Hilbert spaces of smooth and generalized functions. Consider the convolution
\begin{equation}
\label{alpha}
(f\ast\rho)(x)=\int f(y)e^{-(x-y)^{2}}dy
\end{equation}
of a function $f \in L_{2}(R)$ and the Gaussian function $\rho(x)=e^{-x^{2}}$.
It is the standard result that such a convolution is in $C^{\infty}\cap L_{2}(R)$ and
\begin{equation}
D^{p}(f\ast\rho)(x)=(f\ast D^{p}\rho)(x),
\end{equation}
for any order $p$ of the derivative $D$. 

{\flushleft {\bf Theorem 2.1} {\em The linear operator $\rho : L_{2}(R) \longrightarrow L_{2}(R)$ defined by 
\begin{equation}
\rho f =f \ast \rho
\end{equation}
is a bounded invertible operator. }
\smallskip

{\it Proof}. 
The operator $\rho$ is bounded because $\left\| \rho \ast f \right\|_{L_{2}} \le \left\| \rho \right\|_{L_{1}}\left\| f \right\|_{L_{2}}$. To check that it is invertible, assume that $(\rho \ast f)(x)=0$ and differentiate both sides of this equation an arbitrary number of times $p$. We then conclude that all Fourier coefficients $c_{p}$ of $f$ in the orthonormal in $L_{2}(R)$ basis of Hermite functions vanish.}
\bigskip

The operator $\rho$ induces  the Hilbert metric on the image $H=\rho (L_{2}(R))$ by
\begin{equation}
\label{rrr}
(\varphi,\psi)_{H}=(\rho^{-1}\varphi, \rho^{-1}\psi)_{L_{2}}.
\end{equation}
In particular, the space $H$ with this metric is Hilbert.

{\flushleft {\bf Theorem 2.2} {\em The embedding $H\subset L_{2}(R)$ is continuous and $H$ is dense in $L_{2}(R)$.}
\smallskip

{\it Proof}.
We have 
\begin{equation}
\left\|\varphi\right\|_{L_{2}}=\left\|\rho f\right\|_{L_{2}}\le \left\| \rho \right\|_{L_{1}}\left\| f \right\|_{L_{2}}= M\left\|\varphi\right\|_{H},
\end{equation}
where $M$ is a constant.
In addition, the functions $e^{-(x-a)^{2}}$, $a \in R$, being in $H$, form a complete system in $L_{2}(R)$. Therefore, $H$ is dense in  $L_{2}(R)$.}
\bigskip

Provided we identify those linear functionals on $H$ and $L_{2}(R)$ which are equal on $H$, we then have
\begin{equation}
L^{\ast}_{2}(R) \subset H^{\ast}
\end{equation}
Indeed, any functional continuous on $L_{2}$ will be continuous on $H$.

{\flushleft {\bf Theorem 2.3} {\em When using the standard norm
\begin{equation}
\left\|f\right\|_{H^{\ast}}={\sup_{\left\| \varphi \right\|_{H}\le 1}}\left |(f,\varphi)\right| 
\end{equation}
on $H^{\ast}$ and similarly on $L^{\ast}_{2}$, we have a continuous embedding $L^{\ast}_{2}(R) \hookrightarrow H^{\ast}$.}

{\it Proof}. Because $\left\|\varphi\right\|_{L_{2}}\le  M\left\|\varphi\right\|_{H}$, for any $f \in L^{\ast}_{2}$ we have
\begin{equation}
\left\|f\right\|_{H^{\ast}}\le N\left\|f\right\|_{L^{\ast}_{2}}.
\end{equation}
That is, we have a continuous embedding $L^{\ast}_{2}(R) \hookrightarrow H^{\ast}$.}
\bigskip

By Riesz' theorem the norm on $H^{\ast}$ is induced by the inner product
\begin{equation}
\label{i}
(f,g)_{H^{\ast}}=(\varphi, \psi)_{H},
\end{equation}
where $f=(\varphi, \cdot)_{H}={\widehat G}\varphi$ and $g=(\psi, \cdot)_{H}={\widehat G}\psi$ and ${\widehat G}: H \longrightarrow H^{\ast}$ is the Riesz' isomorphism. Similarly for $L^{\ast}_{2}$.

On the other hand, since $\rho: L_{2} \longrightarrow H$ is an isomorphism of Hilbert spaces, the adjoint operator $\rho^{\ast}: H^{\ast} \longrightarrow L^{\ast}_{2}$, defined for any $f \in H^{\ast}$, $\varphi \in L_{2}$ by
\begin{equation}
(\rho^{\ast}f, \varphi)=(f, \rho \varphi),
\end{equation}
is continuous and invertible (with $(\rho^{\ast})^{-1}=(\rho^{-1})^{\ast}$). 

{\flushleft {\bf Theorem 2.4} {\em The operator $\rho^{\ast}$ is an isomorphism of Hilbert spaces $H^{\ast}$ and $L^{\ast}_{2}$ with the above inner product (\ref{i}). The metric on $H^{\ast}$ can be defined by the isomorphism ${\widehat G}^{-1}: H^{\ast} \longrightarrow H$, ${\widehat G}^{-1}=\rho \rho^{\ast}$. }
\smallskip

{\it Proof}.
For any $f,g \in H^{\ast}$ the inner product $(f,g)_{H^{\ast}}$ induced by $\rho^{\ast}$ is given by
\begin{equation}
\label{ii}
(f,g)_{H^{\ast}}=(\rho^{\ast}f, \rho^{\ast}g)_{L^{\ast}_{2}}=(f, \rho \rho^{\ast}g),
\end{equation}
where $\rho \rho^{\ast}: H^{\ast} \longrightarrow H$ defines the metric on $H^{\ast}$.  On the other hand, if $f=(\varphi, \cdot)_{H}$ and $g=(\psi, \cdot)_{H}$, then  $f={\widehat G}\varphi$, and $g={\widehat G}\psi$,  where ${\widehat G}:H \longrightarrow H^{\ast}$ is the Riesz' isomorphism. Therefore, metric (\ref{i}) on $H^{\ast}$ can be written as follows:
\begin{equation}
\label{iii}
(f,g)_{H^{\ast}}=(\varphi,\psi)_{H}=({\widehat G}\varphi, \psi)=({\widehat G}{\widehat G}^{-1}f, {\widehat G}^{-1}g)=(f,{\widehat G}^{-1}g).
\end{equation}
From (\ref{rrr}) we conclude that ${\widehat G}^{-1}=\rho \rho^{\ast}$. Also, because the inner products (\ref{ii}) and (\ref{iii}) coincide, $\rho^{\ast}$ is an isomorphism of Hilbert spaces $H^{\ast}$ and $L^{\ast}_{2}$. }
\bigskip

Because the kernel of $\rho$ is given by $\rho(x,y)=e^{-(x-y)^{2}}$, the kernel $k(x,y)$ of ${\widehat G}^{-1}$, computed up to a constant factor, is equal to
\begin{equation}
\label{k}
k(x,y)=e^{-\frac{(x-y)^{2}}{2}}.
\end{equation}
Therefore, the inner product on $H^{\ast}$ can be written as
\begin{equation}
\label{no}
(\varphi,\psi)_{H^{\ast}}=\int e^{-\frac{(x-y)^{2}}{2}}\varphi(x)\psi(y)dxdy.
\end{equation}
Note that the integral symbol here denotes the action of the bilinear functional $G^{-1}$ on $H^{\ast} \times H^{\ast}$. Only in special cases does this action coincides with Lebesgue integration with respect to $x,y$.

{\flushleft {\bf Theorem 2.5} {\em The space $H^{\ast}$ contains the pointwise evaluation functional (the delta-function) $\delta_{a}:\varphi \longrightarrow \varphi(a)$. } 
\smallskip

{\it Proof}. The sequence 
\begin{equation}
f_{L}=\frac{L}{\sqrt{\pi}}e^{-L^{2}x^{2}}
\end{equation}
is fundamental in $H^{\ast}$ and so it converges in $H^{\ast}$. Also, the sequence $f_{L}$ is a delta-converging sequence, so that $(f_{L},\varphi)\longrightarrow (\delta_{a},\varphi)$ as $L\longrightarrow \infty$ for any $\varphi \in H$. In particular, $f_{L}$ strongly converges to the evaluation functional $\delta_{a}$ in $H^{\ast}$.}
\bigskip

Formally,
\begin{equation}
\label{inn}
(\delta_{a},\delta_{b})_{H^{\ast }}=\int e^{-\frac{(x-y)^{2}}{2}}\delta (x-a)\delta
(y-b)dxdy=e^{-\frac{(a-b)^{2}}{2}}.
\end{equation}

Note that the inner product on $L_{2}(R)$ can be written in the form
\begin{equation}
\label{inn_l}
(\varphi, \psi)_{L_{2}}=\int \delta(x-y)\varphi(x)\psi(y)dxdy,
\end{equation}
where $\int \delta(x-y)\psi(y)dy$ denote the convolution $(\delta \ast \psi)(x)=\psi(x)$. The main difference between metrics (\ref{no}) and (\ref{inn_l}) is that the metric on $L_{2}(R)$ is ``diagonal'', while the metric on $H^{\ast}$ is not.

Consider now yet another Hilbert space obtained in a similar fashion.
For this note that since $H\hookrightarrow L_{2}(R)$, the Fourier transform $\sigma$ on $H$ is defined. Moreover, $\sigma$ induces a Hilbert metric on the image ${\widetilde H}=\sigma(H)$ by
\begin{equation}
\label{Hsig}
\left(\sigma(\varphi),\sigma(\psi)\right)_{{\widetilde H}}=(\varphi,\psi)_{H},
\end{equation}
for any $\varphi,\psi$ in $H$.

We can then define the Fourier transform on $H^{\ast}$ in the standard way by
\begin{equation}
\left(\sigma(f),\sigma(\varphi)\right)=2\pi(f,\varphi),
\end{equation}
for any $f \in H^{\ast}$ and $\varphi \in H$. The image $\sigma(H^{\ast})$ is the Hilbert space ${\widetilde H}^{\ast}$ dual to ${\widetilde H}$. The metric on this space is induced by the map $\sigma\rho: L_{2} \longrightarrow {\widetilde H}$:
\begin{equation}
(f,g)_{{\widetilde H}^{\ast}}=\left((\sigma \rho)^{\ast} f, (\sigma \rho)^{\ast} g\right )_{L^{\ast}_{2}}.
\end{equation}
In other words, the metric on ${\widetilde H}^{\ast}$ is given by $\sigma\rho\rho^{\ast}\sigma^{\ast}: {\widetilde H}^{\ast} \longrightarrow {\widetilde H}$. The kernel of this metric is then equal to  
\begin{equation}
{\widetilde k}(x,y)=\frac{1}{\sqrt{2\pi}}e^{-\frac{x^{2}}{2}}\delta(x-y),
\end{equation}
where $\delta(x-y)$ is the kernel of the operator $\delta \ast$ of convolution with the delta function.
Note that because of the weight $e^{-\frac{x^{2}}{2}}$, the space ${\widetilde H}^{\ast}$ contains the plane wave functions $e^{ikx}$. 

So, we have the Hilbert space $H$ of $C^{\infty}$ (in fact, analytic) functions, its dual  $H^{\ast}$ containing singular generalized functions, in particular, $\delta$, $D^{p}{\delta}$. We also have the Fourier transformed space ${\widetilde H}$ and its dual ${\widetilde H}^{\ast}$, which contains, in particular, the plane wave function $e^{ipx}$.

The method used to construct the above spaces of generalized functions is very general. 
All we need is a linear injective map $\rho$ on a space $L_{2}$ of square-integrable functions. 
By changing $\rho$, we change the metric ${\widehat G}$ on the Hilbert space $H=\rho(L_{2})$ and therefore the metric ${\widehat G}^{-1}$ on the dual space $H^{\ast}$. Roughly speaking, by ``smoothing" the kernel of the metric ${\widehat G}^{-1}$, we extend the class of (generalized) functions for which the norm defined by this metric is finite. The same is true if we ``improve" the behavior of the metric for large $|x|$. 
Conversely, by ``spoiling the metric" we make the corresponding Hilbert space ``poor" in terms of a variety of the elements of the space. This is further illustrated by the following theorem.
{\flushleft {\bf Theorem 2.6} {\em If $\rho$ is the map on $L_{2}(R)$ with the kernel $\rho(x,y)=e^{-(x-y)^{2}-x^{2}}$, then the Hilbert space $\rho\left(L_{2}(R)\right)$ with the induced metric is continuously embedded into the Schwartz space $W$ of $C^{\infty}$ rapidly decreasing functions on $R$. }
\smallskip

{\it Proof.}
From {\em Theorem 2.1} we immediately conclude that the map $\rho$ is injective and the image $\rho\left(L_{2}(R)\right)$ consists of $C^{\infty}$ functions. Moreover, we have
\begin{equation}
\left | \int f(y)e^{-(x-y)^{2}-x^{2}}dy\right| \le M \left\|f\right\|_{L_{2}}e^{-x^{2}},
\end{equation}
where $M$ is a constant.
In particular, the functions in $\rho\left(L_{2}(R)\right)$ decrease faster than any power of $1/|x|$. 

Now, the topology on $W$ may be defined by the countable system of
norms
\begin{equation}
\left\| \varphi \right\| _{p}={\sup_{x \in R;k,q\leq p}}\left|
x^{k}\varphi ^{(q)}(x)\right| ,
\end{equation}
where $k,q,p$ are nonnegative integers and $\varphi ^{(q)}$ is the
derivative of $\varphi $ of order $q$. For any $\varphi \in H$ by the Schwarz inequality we have:
\begin{eqnarray}
\left|x^{k}\varphi ^{(q)}(x)\right|=\left|\int x^{k}f(y)\frac{d^{q}}{dx^{q}}e^{-(x-y)^{2}-x^{2}}dy\right|\ \nonumber \\ 
\leq \left(\int x^{2k}P_{q}^{2}(x)e^{-2(x-y)^{2}-2x^{2}}dy\right)^{\frac{1}{2}}\left\| f \right\| _{L_{2}}=
M_{k,q}\left\| f \right\|_{L_{2}},
\end{eqnarray}
where $\rho f=\varphi$, $P_{q}$ is a polynomial of degree $q$ and $M_{k,q}$ are constants depending only on $k,q$. This proves that topology on $H$ is stronger then topology on $W$, i.e. $H \subset W$ is a continuous embedding. }
\bigskip

It follows from the theorem that $W^{\ast} \subset H^{\ast}$ as a set. In fact, any functional continuous on $W$ will be continuous on $H$. Moreover, by choosing the strong topology on $W^{\ast}$ one can show that the embedding of $W^{\ast}$ into $H^{\ast}$ is continuous. Notice also that one could choose the weak topology on $W^{\ast}$ instead as the weak and strong topologies on $W^{\ast}$ are equivalent \cite{Gelfand1}.

{\flushleft {\bf Remarks} }
1. \quad The considered Hilbert spaces of generalized functions could have been also obtained by completing spaces of ordinary functions with respect to the discussed inner products. More generally, in many cases the expression
\begin{equation}
(f,g)_{k}=\int k(x,y)f(x)g(y)dxdy,
\end{equation}
where $k$ is a continuous function on a region $D$ in $R^{2n}$ defines an inner product on $L_{2}(R^{n})$. Then the completion with respect to the corresponding metric yields a Hilbert space containing singular generalized functions.

2. \quad The isomorphism $\rho$ defined on $L_{2}(R)$ by $(\rho f)(x)=\int e^{-(x-y)^{2}}f(y)dy$ is closely related to the Segal-Bargmann transform $C_{t}:L_{2}(R^{d}) \longrightarrow H(C^{d})$, where $H(C^{d})$ is the space of holomorphic functions on $C^{d}$, given by
\begin{equation}
(C_{t}f)(z)=\int (2\pi t)^{-\frac{d}{2}}e^{-\frac{(z-y)^{2}}{2t}}f(y)dy.
\end{equation}
The theorem by Segal-Bargmann \cite{Barg} (see also \cite{Hall}) says the following.

{\flushleft {\bf Theorem 2.7} {\em For each $t$ the map $C_{t}$ is an isomorphism of $L_{2}(R^{d})$ onto $HL_{2}(C^{d},\nu _{t})$. Here $HL_{2}(C^{d},\nu _{t})$ denotes the space of holomorphic functions that are square-integrable with respect to the measure $\nu_{t}(z)dz$, where $dz$ is the $2d$-dimensional Lebesgue measure on $C^{d}$ and the density $\nu_{t}$ is given by 
\begin{equation}
\nu_{t}(x+iy)=(\pi t)^{-\frac{d}{2}} e^{-\frac{y^{2}}{t}}, \quad x,y \in R^{d}.
\end{equation}
}
}
\smallskip

3. \quad Let us also remark that Hilbert spaces on which the action of pointwise evaluation functional is continuous are often called {\em reproducing kernel Hilbert spaces}.

\section{Generalized eigenvalue problem and functional tensor equations}
\setcounter{equation}{0}

Consider the {\em generalized eigenvalue problem} for a linear operator ${\widehat A}$ on a Hilbert space $H$:
\begin{equation}
\label{eigen_new}
f({\widehat A}\varphi )=\lambda f(\varphi ).
\end{equation}
The problem consists in finding all functionals $f \in H^{\ast}$ and the corresponding numbers $\lambda$ for which the equation (\ref{eigen_new}) is satisfied for all functions $\varphi$ in a Hilbert space $H$ of functions.

For instance, let ${\widetilde H}$ be the Hilbert space with the metric (\ref{Hsig}). As we know, the metric on the dual space ${\widetilde H}^{\ast}$ is given by the kernel $e^{-\frac{x^{2}}{2}}\delta(x-y)$.
The generalized eigenvalue problem for the operator ${\widehat A}=-i\frac{d}{dx}$ is given by
\begin{equation}
\label{eigenvalue1}
f\left (-i\frac{d}{dx}\varphi \right )=p f\left (\varphi \right ).
\end{equation}
The equation (\ref{eigenvalue1}) must be satisfied for every $\varphi$ in ${\widetilde H}$. The functionals
\begin{equation}
f(x)=e^{ip x}
\end{equation}
are the eigenvectors of $A$. Note that these eigenvectors belong to ${\widetilde H}^{\ast}$ so that the generalized eigenvalue problem has a solution. 

As before, the Fourier transform $\sigma:
\widetilde{H}\longrightarrow H$,
\begin{equation}
\psi (k)=(\sigma \varphi )(k)=\int \varphi (x)e^{ikx}dx,
\end{equation}
induces a Hilbert structure on the space $H=\sigma({\widetilde H})$. Relative to this structure $\sigma$ is an isomorphism of the Hilbert spaces $\widetilde{H}$ and $H$. The inverse transform $\omega=\sigma^{-1}$
is given by
\begin{equation}
(\omega \psi )(x)=\frac{1}{2\pi }\int \psi (k)e^{-ikx}dk.
\end{equation}
Notice that the Fourier transform of $e^{ipx}$ is $\delta (k-p)$ which is to say again that the space $H^{\ast}$ dual to $H$ contains delta-functions. As we know from (\ref{k}), the kernel of the metric on $H^{\ast}$ is proportional to $e^{-\frac{1}{2}(x-y)^{2}}$.

Using transformations $\sigma, \omega$ we can rewrite the generalized eigenvalue problem (\ref{eigenvalue1}) as
\begin{equation}
\omega ^{\ast }f(\sigma {\widehat A}\omega \psi )=p \omega ^{\ast }f(\psi ),
\end{equation}
where $\omega \psi=\varphi$. We have:
\begin{equation}
{\widehat A}\omega \psi =-i\frac{d}{dx}\frac{1}{2\pi }\int \psi (k)e^{-ikx}dk=\frac{1}{
2\pi }\int k\psi (k)e^{-ikx}dk.
\end{equation}
Therefore,
\begin{equation}
(\sigma {\widehat A}\omega \psi )(k)=k\psi (k).
\end{equation}
So, the new eigenvalue problem is as follows:
\begin{equation}
\label{eigenvalue2}
g(k\psi )=p g(\psi ).
\end{equation}
Thus, we have the eigenvalue problem for the operator of multiplication by
the variable. The eigenfunctions here are the elements of $H^{\ast}$ given by
\begin{equation}
g(k)=\delta (p-k),
\end{equation}
where $g=\omega ^{\ast }f$. Indeed,
\begin{equation}
(\omega ^{\ast }f)(k)=\frac{1}{2\pi }\int f(x)e^{-ikx}dx=\frac{1}{2\pi }\int
e^{ip x}e^{-ikx}dx=\delta (p-k ).
\end{equation}

Two important things  are happening here. First of all, for the given operator ${\widehat A}$ we have been able to find an appropriate Hilbert space of generalized functions so that the generalized eigenvectors of the operator are its elements. We also conclude that each eigenvalue problem actually defines a whole family of ``unitary equivalent'' problems obtained via isomorphisms of Hilbert spaces.

This is analogous to the case of a single tensor equation on $R^{n}$ considered in different linear coordinates (i.e. in different bases). Instead of a change in {\em numeric} components of vectors and other tensors we now have a change in functions, operators on spaces of functions etc.

The idea is then to consider different equivalence classes of functional objects related by isomorphisms of Hilbert spaces as realizations of invariant objects, which themselves are independent of any particular functional realization.
In particular, the eigenvalue problems (\ref{eigenvalue1}), and (\ref{eigenvalue2}) can be considered as two coordinate expressions of a single equation which is itself independent of a chosen realization. In applications, such realizations, although mathematically equivalent, may represent different physical situations. In the next section this approach will be formalized.

\section{Functional coordinate formalism on Hilbert manifolds}
\setcounter{equation}{0}

Here are the main definitions:
{\flushleft A {\em string space} 
$\bf {S}$
is an abstract topological vector space topologically linearly isomorphic to a separable Hilbert space.}

{\flushleft The elements $\Phi,\Psi,...\quad$ of $\bf {S}$ are called strings. }

{\flushleft A {\em Hilbert space of functions} (or a {\em coordinate space}) is either a Hilbert space
$H$, elements of which are equivalence classes of maps between two given
subsets of $R^{n}$ or the Hilbert space $H^{\ast }$ dual to $H$. }

{\flushleft A linear isomorphism ${e}_{H}$ from a Hilbert space $H$
of functions onto $\bf {S}$
is called a {\em string basis} (or a {\em functional basis}) on $\bf {S}$.}

{\flushleft The inverse map ${e}^{-1}_{H}: {\bf S} \longrightarrow H$ is called a {\em linear coordinate system on ${\bf S}$} (or a {\em linear functional coordinate system}).}

{\flushleft {\bf Analogy}:
${\bf S}$ is analogous to the abstract $n$-dimensional vector space $V$.
A coordinate space $H$ is analogous to a particular realization $R^{n}$ of $V$ as a space of column vectors.
A string basis $e_{H}$ is analogous to the ordinary basis $\{e_{k}\}$ on $V$. 
Namely, the string basis identifies a string with a function: if $\Phi \in {\bf S}$, then ${\Phi }={e}_{H}(\varphi)$ for a unique $\varphi \in H$.
The ordinary basis $\{e_{k}\}$ identifies a vector with a column of numbers. It can be therefore identified with an isomorphism $e_{k}:R^{n} \longrightarrow V$. The inverse map $e^{-1}_{k}$ is then a global coordinate chart on $V$ or a linear coordinate system on $V$. }

{\flushleft {\bf Remark.} Of course, any separable Hilbert space ${\bf S}$ can  be realized as a space $l_{2}$ of sequences by means of an orthonormal basis on ${\bf S}$. So it seems that we already have a clear infinite-dimensional analogue of the $n$-dimensional vector space. However, such a realization is too restrictive. In fact, unlike the case of a finite number of dimensions, there exist various realizations of ${\bf S}$ as a space of functions. These realizations are also ``numeric'' and useful in applications and they should not be discarded. Moreover, we will see that they can be put on an equal footing with the space $l_{2}$. }
\bigskip
 
The basis ${e}_{H^{\ast }}:H^{\ast} \longrightarrow {\bf {S}}^{\ast}$ is called {\em dual} to the
basis ${e}_{H}$ if for any string ${\Phi }={e}_{H}(\varphi)$ and for any
functional ${F}={e}_{H^{\ast }}(f)$ in ${\bf {S}}^{\ast}$ the following is true:
\begin{equation}
{F}({\Phi })=f(\varphi ).
\end{equation}

{\flushleft {\bf Analogy}:
Dual string basis ${e}_{H^{\ast }}$ is analogous to the ordinary 
basis $\{e^{m}\}$ dual to a given basis $\{e_{k}\}$. Indeed, if $\Phi$ is a vector in the finite dimensional vector space $V$, and $F \in V^{\ast}$ is a linear functional on $V$, then
\begin{equation}
F(\Phi)=f^{k}\varphi_{k},
\end{equation}
where $f^{k}$ and $\varphi_{k}$ are components of $F$ and $\Phi$ in the bases $\{e^{m}\}$ and $\{e_{k}\}$ respectively.}

\bigskip
By definition the string space $\bf {S}$ is isomorphic to a
separable Hilbert space. Assume that $\bf {S}$ itself is a Hilbert space. 
Assume further that the string bases
$e_{H}$ are isomorphisms of Hilbert spaces. That is,
the Hilbert metric on any coordinate space $H$ is determined by the Hilbert metric on $\bf {S}$ and the choice of a string basis. Conversely, 
{\flushleft {\bf Proposition 4.1 } {\em The choice of a coordinate Hilbert space determines the
corresponding string basis $e_{H}$ up to a unitary transformation.}
\smallskip

{\it Proof.} Indeed, with $H$ fixed, any two bases $e_{H}$, 
${\widetilde e}_{H}$ can only differ by an automorphism of $H$, i.e., by a unitary transformation. }
\bigskip

Assume for simplicity that $H$ is a real Hilbert space (generalization to the case of a complex Hilbert space will be obvious). We have:
\begin{equation}
\label{orto}
(\Phi ,\Psi )_{S}={\bf{G}}(\Phi ,\Psi )=G(\varphi,\psi)=\int g(x,y)\varphi(x)\psi(y)dxdy
=g_{xy}\varphi ^{x}\psi ^{y},
\end{equation}
where ${\bf {G}}:{\bf {S}}\times {\bf {S}}\longrightarrow R$
is a bilinear functional defining the inner
product on $\bf {S}$ and $G:H\times H\longrightarrow R$ is the induced bilinear functional. As before, the integral sign is a symbol of action of $G$ on $H\times H$ and the expression on the right is a convenient
form of writing this action. 

A string basis ${e}_{H}$ in $\bf {S}$ will be called
{\em orthogonal} if for any $\Phi, \Psi \in {\bf S}$ we have
\begin{equation}
(\Phi ,\Psi )_{S}=f_{\varphi}(\psi ),
\end{equation}
where $f_{\varphi}$ is a {\it regular} functional and $\Phi=e_{H}\varphi$,
$\Psi=e_{H}\psi$ as before. That is,
\begin{equation}
\label{orto1}
(\Phi,\Psi)_{S}=f_{\varphi}(\psi)=\int \varphi(x)\psi(x)d\mu (x),
\end{equation}
where $\int$ here denotes an actual integral over a $\mu $-measurable set $D \in R^{n}$ which is the domain of definition of functions in $H$.

If the integral in (\ref{orto1}) is the usual Lebesgue integral and/or a sum over a discrete index $x$, the corresponding coordinate space will be called an $L_{2}$-space. In this case we will also say that the basis $e_{H}$ is {\em orthonormal}.
If the integral is a more general Lebesgue-Stieltjes integral, the coordinate space defined by (\ref{orto1}) will be called an $L_{2}$-space with the weight $\mu$ and the basis $e_{H}$ will be called {\em orthogonal}.

{\flushleft {\bf Analogy}:
An orthonormal string basis is analogous to the usual orthonormal basis. In fact, if $\varphi^{k}$, $\psi^{k}$  are components of vectors $\Phi, \Psi$ in an orthonormal basis $e_{k}$ on an Euclidean space $V$, then $(\Phi,\Psi)_{V}=\delta_{kl}\varphi^{k}\psi^{l}=\varphi^{k}\psi^{k}$, which is similar to (\ref{orto1}).}

Roughly speaking, the metric on Hilbert spaces defined by orthogonal string bases has a ``diagonal'' kernel. In particular, the kernel may be proportional to the delta-function (as in $L_{2}(R)$) or to the Kr{\oe}necker symbol (as in $l_{2}$). More general coordinate Hilbert spaces have a ``non-diagonal'' metric. The metrics (\ref{no}) and (\ref{inn_l}) provide examples.

It is important to distinguish clearly the notion of a string basis from the notion of an ordinary basis on a Hilbert space of functions.
Namely, a  string basis permits us to represent invariant objects in string space (strings) in terms of functions, which are elements of a Hilbert space of functions. A basis on the space of functions then allows us
to represent functions in terms of numbers; that is, in terms of the components
of the functions in the basis. On the other hand, we have the following proposition:

{\flushleft {\bf Proposition 4.2} {\em The ordinary orthonormal basis on ${\bf S}$ can be identified with the string basis $e_{l_{2}}:l_{2} \longrightarrow {\bf S}$.  }
\smallskip

{\it Proof.}
The ordinary orthonormal basis on ${\bf S}$ provides an isomorphism $\omega$ from ${\bf S}$ onto $l_{2}$ which associates with each element $\Phi$ in ${\bf S}$ the sequence of its Fourier components in the basis. Conversely, given an isomorphism $\omega: {\bf S} \longrightarrow l_{2}$, we can associate with each $\Phi \in {\bf S}$ the sequence $\omega(\Phi)$ of its Fourier components in a unique (for all $\Phi$) orthonormal basis. Therefore, the ordinary orthonormal basis on ${\bf S}$ can be identified with the string basis $e_{l_{2}}=\omega^{-1}$. }
\bigskip

A {\em linear coordinate transformation on $\bf {S}$} is
an isomorphism $\omega :\widetilde{H}\longrightarrow H$ of Hilbert
spaces which defines a new string
basis $e_{\widetilde{H}}:\widetilde{H}\longrightarrow \bf {S}$ by
$e_{\widetilde{H}}=e_{H}\circ \omega $.

{\flushleft {\bf Proposition 4.3}
{\em Let $\varphi=e^{-1}_{H}\Phi $, ${\widehat A}=e^{-1}_{H}{\bf {\widehat A}}e_{H}$ and ${\widehat G}=\left( e^{-1}_{H}\right)^{\ast}{\bf {\widehat G}}e^{-1}_{H}$ be the coordinate expressions of a string $\Phi$, an operator ${\bf {\widehat A}}: {\bf S} \longrightarrow {\bf S}$ and the metric $\widehat{\bf{G}}:{\bf {S}}
\longrightarrow {\bf{{S}}^{\ast }}$ in a basis $e_{H}$. Let $\omega :\widetilde{H}\longrightarrow H$ be a linear coordinate transformation on ${\bf S}$.
Then we have the following transformation laws:
\begin{eqnarray}
\label{tensor}
\varphi &=&\omega\widetilde{\varphi}\\
\label{tensor1}
\widehat{G}_{\widetilde{H}} &=&\omega^{*}\widehat{G}\omega\\
\label{tensor3}
{\widehat A}_{\widetilde{H}} &=&\omega^{-1}{\widehat A}\omega,
\end{eqnarray}
where $\widetilde{\varphi}$, ${\widehat A}_{\widetilde{H}}$ and $\widehat{G}_{\widetilde{H}}$ are coordinate functions of $\Phi$, ${\bf {\widehat A}}$ and $\widehat{\bf{G}}$ in the basis $e_{\widetilde{H}}$.}
\smallskip

{\it Proof.} This follows immediately from 
\begin{equation}
(\widehat{G}\varphi,{\widehat A}\psi)=
(\omega^{*}\widehat{G}\omega\widetilde{\varphi},\omega^{-1}{\widehat A}\omega\widetilde{\psi})=
(\widehat{G}_{\widetilde{H}}\widetilde{\varphi},{\widehat A}_{\widetilde{H}}\widetilde{\psi}).
\end{equation}
}
\bigskip

The correspondence between the classical and ``string'' definitions can be summarized in the table:
\bigskip

\begin{table}[ht]
\begin{tabular}{||l|l|l||}\hline
%$e_{H}$ & $e_{k}$ \\\hline
basis & $e_{H}:H \rightarrow {\bf S}$ & $e_{k}:R^{n}\rightarrow V$ \\ \hline 
inner product & $(\Phi,\Psi)_{{\bf S}}=\int g(x,y)\varphi(x)\psi(y)dxdy$ & $(\Phi,\Psi)_{V}=g_{kl}\varphi^{k}\psi^{l}$ \\\hline
dual basis & $e_{H^{\ast}}$, $F(\Phi)=f(\varphi)$ &  $e^{k}$,  $F(\Phi)=f_{k}\varphi^{k}$ \\\hline
ortho basis & $e_{L_{2}}$, $(\Phi,\Psi)_{{\bf S}}=\int \varphi(x)\psi(x)dx$ &  $e_{k}$, $(\Phi,\Psi)_{V}=\delta_{kl}\varphi^{k}\psi^{l}$\\\hline
change of basis & $e_{\widetilde{H}}=e_{H}\circ \omega $, $\widehat{G}_{\widetilde{H}}= \omega^{\ast}\widehat{G}\omega$,...& $e_{{\widetilde k}}=e_{k}\omega_{\widetilde k}^{k} $, $\widehat{G}_{{\widetilde k} {\widetilde l}}=$...\\\hline
\end{tabular}
\caption{Classical versus string bases}
\end{table}
\bigskip

More generally, consider an arbitrary Hilbert manifold ${\it S}$ modeled on ${\bf S}$.
Let $(U_{\alpha },\pi _{\alpha })$ be an atlas on $\it {S}$.
A collection of quadruples $(U_{\alpha },\pi _{\alpha
},\omega _{\alpha },H_{\alpha })$, where each $H_{\alpha }$ is a Hilbert
space of functions and $\omega _{\alpha }$ is an isomorphism of $\bf {S}$
onto $H_{\alpha }$ is called a {\em functional atlas} on
$\it {S}$. A collection of all compatible (i.e., related by diffeomorphisms) functional atlases on $\bf {S}
$ is called a {\em coordinate structure} on  $\it {S}$. A Hilbert manifold ${\it S}$ with the above coordinate structure is called a {\em string manifold} or a {\em functional manifold}.

Let $(U_{\alpha },\pi _{\alpha })$ be a chart on $\it {S}$.
If $p\in U_{\alpha },$ then
$\omega _{\alpha }\circ \pi _{\alpha }(p)$ is called the {\em coordinate}
of $p$. The map $\omega _{\alpha }\circ \pi _{\alpha }:U_{\alpha} \longrightarrow H_{\alpha}$ is called a {\em coordinate system}. The diffeomorphisms
$\omega _{\beta }\circ \pi _{\beta }\circ (\omega _{\alpha }\circ
\pi _{\alpha })^{-1}:\omega _{\alpha }\circ \pi _{\alpha }
(U_{\alpha }\cap U_{\beta })\longrightarrow
\omega _{\beta }\circ \pi _{\beta }(U_{\alpha }\cap U_{\beta })$ are called
{\em string (or functional) coordinate transformations}.

As $\it {S}$ is a differentiable manifold one can also introduce the tangent
bundle structure $\tau :T{\it{S}}\longrightarrow {\it {S}}$ and the bundle $\tau
_{s}^{r}:T_{s}^{r}{\it {S}}\longrightarrow {\it {S}}$
of tensors of rank $(r,s)$. Whenever necessary to distinguish tensors (tensor fields) on ordinary Hilbert manifolds from tensors on manifolds with a coordinate structure, we will call the latter tensors the {\em string tensors} or the {\em functional tensors}. Accordingly, the equations invariant under string coordinate transformations will be called the {\em string tensor} or the {\em functional tensor equations}. 

A coordinate structure on a Hilbert manifold permits one to obtain a functional
description of any string tensor. Namely, let ${\bf{G}}_{p}(F_{1},...,F_{r},\Phi
_{1},...,\Phi _{s})$ be an $(r,s)$-tensor on $\it {S}$.
The coordinate map $\omega _{\alpha }\circ \pi _{\alpha
}:U_{\alpha }\longrightarrow H_{\alpha }$ for each $p\in U_{\alpha }$\ yields
the linear map of tangent spaces $d\rho _{\alpha }:T_{\omega _{\alpha }\circ
\pi _{\alpha }(p)}H_{\alpha }\longrightarrow T_{p}\it {S}$, where $\rho
_{\alpha }=\pi _{\alpha }^{-1}\circ \omega _{\alpha }^{-1}.$ This map is
called a {\it local coordinate string basis on} $\it {S}$.
Notice that for each $p$ the map\ $e_{H_{\alpha }} \equiv e_{H_{\alpha }}(p)$ is a string basis as
defined earlier. Therefore, the local dual basis $e_{H_{\alpha
}^{\ast }}=e_{H_{\alpha }^{\ast }}(p)$ is defined for each $p$
as before and is a function of $p.$

We now have $F_{i}=e_{H_{\alpha }^{\ast }}f_{i}$, and $\Phi
_{j}=e_{H_{\alpha }}\varphi _{j}$ for any $F_{i}\in T_{p}^{\ast }\it {S}$
, $\Phi _{j}\in T_{p}\it {S}$ and some $f_{i}\in H_{\alpha }^{\ast
},\varphi _{j}\in H_{\alpha }$. Therefore the equation
\begin{equation}
{\bf{G}}_{p}(F_{1},...,F_{r},
\Phi _{1},...,\Phi _{s})=G_{p}(f_{1},...,f_{r},\varphi _{1},...,\varphi _{s})
\end{equation}
defines component functions of the $(r,s)$-tensor ${\bf{G}}_{p}$ in the
local coordinate basis $e_{H_{\alpha }}$.

There are two important aspects associated with the formalism. First of all, by changing the string basis $e_{H}$ we can reformulate a given problem on Hilbert space in any way we want. In particular, we can go back and forth between spaces of smooth and generalized functions and choose the one in which the problem in hand is simpler to deal with. In this respect the theory of generalized functions is now a part of the above formalism on Hilbert spaces. 

The second aspect is that we can now work directly with families of ``unitary equivalent'' problems rather than with particular realizations only. In fact, a single string tensor equation describes infinitely many equations on various spaces of functions.

In the following sections various examples and related propositions and theorems will illustrate the importance of these two aspects. 
%%%%%%%%%%%%%%%%%%%%%%

%Once again, the above coordinate formalism deals with a different ``quality" of elements than its finite dimensional counterpart. By changing a string basis we change such properties of coordinates of a string as smoothness, convergence, etc. These properties make no sense for the finite dimensional coordinates of a point which simply are columns of numbers.

\section{The generalized eigenvalue problem as a functional tensor equation}
\setcounter{equation}{0}

Let us begin with a {\em string basis independent} formulation of the discussed earlier generalized eigenvalue problem.
Namely, consider the generalized eigenvalue problem 
\begin{equation}
\label{LLLL}
F({\bf{\widehat A}}\Phi)=\lambda F(\Phi),
\end{equation}
for a linear operator ${\bf {\widehat  A}}$ on ${\bf S}$. The problem consists in finding all functionals $F \in {\bf S}^{\ast}$ and the corresponding numbers $\lambda$ for which the string tensor equation (\ref{LLLL}) is satisfied for all $\Phi \in {\bf S}$.

Assume that the pair $F, \lambda$ is a solution of (\ref{LLLL}) and $e_{H}$ is a string basis on ${\bf S}$. Then we have
\begin{equation}
\label{1111}
e_{H}^{\ast }F(e_{H}^{-1}{\bf{\widehat A}}e_{H}\varphi )=\lambda e_{H}^{\ast
}F(\varphi ),
\end{equation}
where $e_{H}\varphi =\Phi $ and
$e_{H}^{-1}{\bf{\widehat A}}e_{H}$ is the representation of ${\bf{\widehat A}}$ in
the basis $e_{H}$.
By defining $e_{H}^{\ast }F=f$ and ${\widehat A}=e_{H}^{-1}{\bf{\widehat A}}e_{H}$, we have
\begin{equation}
\label{eigen_new1}
f({\widehat A}\varphi )=\lambda f(\varphi ).
\end{equation}
The latter equation describes not just one eigenvalue problem, but
a family of such problems, one for each string basis $e_{H}$. As we change
$e_{H}$, the operator ${\widehat A}$ in general changes as
well, as do the eigenfunctions $f$. 

Assume that in a particular string basis $e_{H}$ the problem (\ref{eigen_new1}) is the already discussed generalized eigenvalue problem for the operator of differentiation $-i\frac{d}{dx}$. Then (\ref{LLLL}) is nothing but the corresponding ``realization independent'' generalized eigenvalue problem given by a functional tensor equation. 

\section{The spectral theorem}
\setcounter{equation}{0}

Here we apply the coordinate formalism to reformulate the standard results of linear algebra in Hilbert spaces.
{\flushleft {\bf Definition 6.1} A string basis $e_{H}$ is called the {\it proper basis} of
a linear operator ${\bf{A}}$ on $\bf {S}$ with eigenvalues (eigenvalue function)
$\lambda=\lambda(k)$, if
\begin{equation}
\label{eb0}
{\bf {A}}e_{H}(\varphi)=e_{H}(\lambda\varphi)
\end{equation}
for any $\varphi \in H$.}
\bigskip

As any string basis, the proper basis of $\bf {A}$ is a linear map from
$H$ onto $\bf {S}$ and a numeric function $\lambda$ is defined on the same set as functions $\varphi \in H$.

By rewriting (\ref{eb0}) as
\begin{equation}
{e_{H}}^{-1}{\bf {A}}e_{H}(\varphi)=\lambda\varphi
\end{equation}
we see that the problem of finding a proper basis of $\bf {A}$ is equivalent to
the problem of finding such a string basis $e_{H}$ in which the action of $\bf {A}$
reduces to multiplication by a function $\lambda$. 

Because the expression ``the adjoint of an operator'' has at least two different meanings, let's accept the following:

{\flushleft
{\bf Definition 6.2} Let ${\widehat A}$ be a continuous linear operator which maps a space $H$ into a space $\widetilde{H}$. Then the {\it adjoint} ${\widehat A}^{\ast}$ of operator ${\widehat A}$ maps the space $\widetilde{H}^{\ast}$ into the space $H^{\ast}$ according to
\begin{equation}
\label{adjoint}
({\widehat A}^{\ast}f,\varphi)=(f,{\widehat A}\varphi)
\end{equation}
for any $\varphi \in H, f \in \widetilde{H}^{\ast}$. Assume further that $H$ is continuously embedded into an $L_{2}$-space and under the identification $L^{\ast}_{2}=L_{2}$ the action of ${\widehat A}^{\ast}$ and ${\widehat A}$ on their common domain $H$ coincide. Then the operator ${\widehat A}$ will be called {\em self-adjoint}.
}
\bigskip

{\flushleft
{\bf Definition 6.3} Let $A$ be a continuous linear operator on a Hilbert space $H$. Then the {\it Hermitian conjugate} operator $A^{+}$ of $A$ is defined on $H$ by
\begin{equation}
\label{hermitian}
(A^{+}\varphi,\psi)_{H}=(\varphi, A\psi)_{H},
\end{equation}
for any $\varphi, \psi \in H$. If $A=A^{+}$, the operator is called Hermitian.
}
\bigskip

Let now $A$ be given on $H$ and let $\widehat{G}: H \longrightarrow H^{\ast }$
define a metric on $H$. Then
\begin{equation}
\label{1}
(\widehat{G}\varphi, A\psi)=(A^{\ast}\widehat{G}\varphi, \psi)=(\widehat{G}A^{+}\varphi, \psi)
\end{equation}
and the relationship between the operators is as follows:
\begin{equation}
\label{2}
A^{+}={\widehat{G}}^{-1}A^{*}{\widehat{G}}.
\end{equation}

{\flushleft {\bf Theorem 6.4} {\em Any Hermitian operator ${\bf {\widehat A}}$ on ${\bf S}$ possesses a proper basis with the eigenvalue function $\lambda(x)=x$.}
\smallskip

{\it Proof}. The theorem simply says that any Hermitian operator is unitary equivalent to the operator of multiplication by the variable, which is the standard result of spectral theorem. } 
\bigskip

%%%%%%%%%%%%%%%%%%%%
%%%%%%%%%%%%%%%%%%%
{\flushleft {\bf Theorem 6.5} {\em Let ${\bf A}$ be an Hermitian operator  on ${\bf S}$ and let $e_{H}$  be a proper basis of ${\bf {\widehat A}}$. Assume that $H$ here is a Hilbert space of numeric valued functions on a (possibly infinite) interval $(a,b)$ and the eigenvalue function is given by $\lambda(x)=x$, $x \in (a,b)$. Assume further that the fundamental space $K$ of infinitely differentiable functions of bounded support in $(a,b)$ is continuously embedded into $H$ as a dense subset. Then the proper basis of ${\bf {\widehat A}}$ is orthogonal.}
\smallskip

{\it Proof}.} In the proper basis $e_{H}$ of ${\bf {\widehat A}}$ we have
\begin{equation}
(\Phi,{\bf A}\Psi)_{S}=(e_{H} \varphi, {\bf A} e_{H}\psi)_{S}=(\varphi,\lambda \psi)_{H}=\int g(x,y)\varphi(x)y\psi(y)dxdy.
\end{equation}
In agreement with section 4, the integral symbol is used here for the action of the bilinear metric functional $G$ with the kernel $g$.
Hermicity gives then
\begin{equation}
\label{9_10}
\int g(x,y)(x-y)\varphi(x)\psi(y)dxdy=0
\end{equation}
for any $\varphi,\psi \in H$. 

Let us show that $g(x,y)=a(x)\delta(x-y)$, i.e. $H$ must coincide with the space $L_{2}(a,b)$ with weight $a(x)$. Note first of all that because $K \hookrightarrow H$, any bilinear functional on $H$ is also a bilinear functional on $K$. 
Also, by the kernel theorem \cite{Gel-Kos} any bilinear functional $G(\varphi, \psi)=\int g(x,y)\varphi (x) \psi (y)dxdy$ on the space $K$ of infinitely differentiable functions of bounded support has the form
\begin{equation}
\label{who}
G(\varphi, \psi)=(f, \varphi (x)\psi (y)),
\end{equation}
where $f$ is a linear functional on the space $K_{2}$ of infinitely differentiable functions $\varphi (x,y)$ of bounded support in $(a,b)\times (a,b) \subset R^{2}$.
If $\Omega \subset R^{2}$ is a bounded domain, $\varphi, \psi \in K$ are arbitrary with support in $\Omega$ and $x \neq y$ on $\Omega$, then
linear combinations of the functions $(y-x)\varphi (x) \psi (y)$ form a dense subset in the space $K_{2}(\Omega)$ of infinitely differentiable functions with support in $\Omega$.
From (\ref{9_10}) it follows then that the functional $f$ in (\ref{who}) is zero on $K_{2}(\Omega)$ for any such $\Omega$. Therefore, the bilinear functional $G$ is concentrated on the diagonal $x=y$ in $R^{2}$. That is, $G$ is equal to zero on all pairs of functions $\varphi, \psi \in H$ such that $\varphi(x)\psi(y)$ is equal to zero in a neighborhood of the diagonal.
Accordingly, the left hand side of (\ref{9_10}) becomes
\begin{equation}
\label{222_1}
\int \sum^{n}_{k=0}a_{k}(x)D^{k}\delta(y-x) (y-x)\varphi(x)\psi(y)dxdy,
\end{equation}
where $D^{k}$ is the derivative of order $k$ with respect to $y$ and $n$ is finite (see \cite{Gelfand1}).

``Integration by parts" in (\ref{222_1}) gives zero for $k=0$ term and
\begin{equation}
\pm \int p\psi^{(p-1)}(x)a_{p}(x)\varphi(x)dx
\end{equation}
for $k=p$ term and any $0<p\leq n$.
By choosing $\psi$ so that the functions $\psi^{(p-1)}, 0<p\leq n$ are linearly independent in $L_{2}(a,b)$, one concludes that the components $a_{p}(x)\varphi(x)$ must be equal to zero almost everywhere in $R$ .  Because $\varphi \in K$ is an arbitrary  function, the functions $a_{p}(x), 0<p\leq n$ must be zero almost everywhere in $R$. Therefore, the functional $G$ is given by the kernel $g(x,y)=a(x)\delta(x-y)$. As $K$ is dense in $H$, this kernel also defines the bilinear functional on $H$.
We conclude that the proper basis of ${\bf {\widehat A}}$ is orthogonal (as defined in section 4).
\bigskip

Note that this theorem is nothing but the generalization to the case of a continuous spectrum of the well known theorem on orthogonality of eigenvectors of an Hermitian operator corresponding to different eigenvalues. 

The conclusion of the theorem seems to favor the use of $L_{2}$-spaces when working with Hermitian operators. Notice however, that in
an orthogonal proper basis of ${\bf {\widehat A}}$ with $H=L_{2}(a,b)$ we have
\begin{equation}
(\Phi,{\bf {\widehat A}}\Psi)_{S}=(f,\lambda g)_{L_{2}}=\int f(x)\lambda(x)g(x)dx,
\end{equation}
where $f,g \in L_{2}(a,b)$. Therefore, it is impossible for $f(x)$ to be a (generalized) eigenvector of the operator ${\widehat A}$ of multiplication by the function $\lambda(x)$.
Indeed, the eigenvectors of such an
operator would  be $\delta$-functionals and the latter ones do not belong to $L_{2}(a,b)$.

To include such functionals in the formalism we need to consider Hilbert spaces containing ``more" functions than $L_{2}(a,b)$.
By the above this in general requires consideration of non-Hermitian operators.

{\flushleft {\bf Example}.} The operator of multiplication by the variable ${\widehat A}=x$ has no eigenvectors in $L_{2}(0,1)$.
Consider then a different Hilbert space $H^{\ast}$ of functions of a real variable $x \in (0,1)$ with the metric $G$ given by a smooth kernel $g(k,m)$ and with the dual space $H$ consisting of continuous functions only. We have:
\begin{equation}
(G\varphi,{\widehat A}\psi)=\int g(k,m)\varphi(k)m\psi(m)dkdm.
\end{equation}
Although the operator ${\widehat A}=x$ is not Hermitian on $H$, it is self-adjoint (see {\em Definition 6.2}). In fact, if $\varphi, \psi$ are continuous functions, then
\begin{equation}
\int g(k,m)\varphi(k)m\psi(m)dkdm = \int mg(k,m)\varphi(k)\psi(m)dkdm
\end{equation}
So on the functionals 
\begin{equation}
F(m)=\int g(k,m)\varphi(k)dk
\end{equation}
which are functions in $H$
we have:
\begin{equation}
F(m\psi)=(mF)(\psi).
\end{equation}
Note also that the eigenfunctions of ${\widehat A}=x$ are in $H^{\ast}$.

The example shows that at least in some cases there is a possibility to accommodate the Hermicity of an operator in $L_{2}$, the self-adjointness of its restriction ${\widehat A}$ onto a Hilbert subspace $H \subset L_{2}$ and the inclusion of generalized eigenvectors of ${\widehat A}$ into the conjugate space $H^{\ast}$. Before formalizing and generalizing this statement we need the following definition (see \cite{Gel-Kos}):

{\flushleft {\bf Definition 6.6} Let $H^{\ast}_{\lambda}$ be the {\it eigenspace} of ${\widehat A}$ consisting of all the generalized eigenvectors $f_{\lambda}$ of ${\widehat A}$ whose eigenvalue is $\lambda$.
Associate with each element $\varphi \in H$ and each number $\lambda$ a linear functional ${\widetilde \varphi}_{\lambda}$ on $H^{\ast}_{\lambda}$ which takes the value $f_{\lambda}(\varphi)$ on the element $f_{\lambda}$.
This gives vector-functions of $\lambda$ whose values are linear functionals on $H^{\ast}_{\lambda}$. The correspondence $\varphi \longrightarrow {\widetilde \varphi}_{\lambda}$ is called the {\it spectral decomposition of $\varphi$ corresponding to the operator ${\widehat A}$}.
The set of generalized eigenvectors of ${\widehat A}$ is called {\it complete} if ${\widetilde \varphi}_{\lambda}=0$ implies $\varphi=0$.}
\bigskip

We have now the following

{\flushleft {\bf Theorem 6.7} {\em Let ${\widehat B}$ be an Hermitian operator on a Hilbert space $L_{2}$. Then there exists a topological Hilbert subspace $H$ of $L_{2}$ which is dense in $L_{2}$ and such that the restriction ${\widehat A}$ of ${\widehat B}$ onto $H$ is a self-adjoint operator and the conjugate space $H^{\ast}$ contains the complete set of eigenvectors of ${\widehat A}$. Moreover, there exists a coordinate transformation $\rho: H \longrightarrow \widetilde {H}$ such that the transformed operator ${\widehat {\widetilde A}}=\rho {\widehat A} \rho^{-1}$ is the operator of multiplication by $x$.}
\smallskip

{\it Proof}.} Notice first of all that the property of an operator ${\widehat A}$ to be Hermitian is invariant under  functional coordinate transformations. 

Let $\omega: L_{2} \longrightarrow L_{2}(R)$ be an isomorphism of Hilbert spaces. By the above ${\widehat B}_{1}= \omega {\widehat B} \omega^{-1}$ is an Hermitian operator in $L_{2}(R)$.

Let $W$ be the Schwartz space of infinitely differentiable functions on $R$. It is known that $W$ is a nuclear space and the triple $W \subset L_{2}(R) \subset W^{\ast}$ is a rigged Hilbert space \cite{Gel-Kos}. The operator ${\widehat B}_{1}$ is an Hermitian operator in the rigged space. From \cite{Gel-Kos} we know that the operator ${\widehat B}_{1}$ has a complete set of generalized eigenvectors belonging to $W^{\ast}$.

Let us choose a Hilbert space $H_{1}$ such that $H_{1}$ is a topological subspace of $W$ which is also dense in $L_{2}(R)$. We know from the {\em Theorem 2.6} that such a Hilbert space $H_{1}$ exists. As $W^{\ast} \subset H^{\ast}_{1}$, the space $H^{\ast}_{1}$ contains the complete set of eigenvectors of the restriction ${\widehat A}_{1}$ of ${\widehat B}_{1}$ onto $H_{1}$. The image $H=\omega^{-1}(H_{1})$ is a Hilbert subspace of $L_{2}$ with the induced Hilbert structure. It is dense in $L_{2}$ and on this subspace the operator ${\widehat A}=\omega^{-1} {\widehat A}_{1}\omega$ is self-adjoint and coincides with the original operator ${\widehat B}$. This proves the first part of the theorem.

Now, by the abstract theorem on spectral decomposition of Hermitian operators there exists a realization $\tau: L_{2}(R)\longrightarrow \widetilde{L}_{2}$ such that the operator ${\widehat B}_{1}$ is given by multiplication by $x$. Here $\widetilde{L}_{2}$ denotes in general a direct integral of Hilbert spaces of the $L_{2}$-type.

Consider the restriction of $\tau$ onto $H_{1}$. As $\tau$ is an isomorphism, it induces a Hilbert structure on the image $\tau(H_{1})=\widetilde{H}$.

Consider then the isomorphism $\kappa =\tau \circ \omega: H \longrightarrow \widetilde{H}$ as a coordinate transformation. This transformation takes the operator ${\widehat A}$ in $H$ into the operator of multiplication by $x$ in $\widetilde{H}$. This completes the proof.
\bigskip

Assume now that ${\widehat A}$ is the operator constructed in the theorem. That is, ${\widehat A}$ is a self-adjoint operator in a Hilbert space $H$ such that the conjugate space $H^{\ast}$ contains the complete set of generalized eigenvectors $f_{\lambda}$ of ${\widehat A}$.
Let $\varphi \longrightarrow {\widetilde \varphi}_{\lambda}$ be the spectral decomposition of the element $\varphi \in H$ corresponding to the operator ${\widehat A}$.
Let $e_{H}$ be a string basis on $\bf {S}$ and $\Phi=e_{H}\varphi, F_{\lambda}=(e^{\ast}_{H})^{-1}f_{\lambda}, {\bf {\widehat A}}=e_{H}{\widehat A}e^{-1}_{H}$ as always.

The spectral decomposition $\varphi \longrightarrow {\widetilde \varphi}_{\lambda}$ establishes the correspondence $\Phi \longrightarrow {\widetilde \varphi}_{\lambda}$ which is the spectral decomposition of the string $\Phi$ corresponding to the operator $\bf {{\widehat A}}$.

More directly, to construct $\Phi \longrightarrow {\widetilde \varphi}_{\lambda}$
we introduce the eigenspace ${\bf S}^{\ast}_{\lambda}$ of $\bf {{\widehat A}}$ which consists of all eigenvectors $F_{\lambda}$ of $\bf {{\widehat A}}$ whose eigenvalue is $\lambda$. Then we associate with each string $\Phi \in {\bf S}$ and each number $\lambda$ a linear functional ${\widetilde \varphi}_{\lambda}$ on ${\bf S}^{\ast}_{\lambda}$ which takes the value $F_{\lambda}(\Phi)$ on the element $F_{\lambda}$.

As the set of generalized eigenvectors $f_{\lambda}$ of ${\widehat A}$ is complete, so is the set of eigenvectors $F_{\lambda}$ of the operator $\bf {{\widehat A}}$. In fact, whenever ${\widetilde \varphi}_{\lambda}=0$ we have $\varphi=0$ by completeness of the set of eigenvectors $f_{\lambda}$. But then $\Phi=0$ as well, i.e. the set of eigenvectors $F_{\lambda}$ of $\bf {{\widehat A}}$ is complete.

The correspondence $\rho_{\widetilde H}: \Phi \longrightarrow {\widetilde \varphi}_{\lambda}$ is an isomorphism of linear spaces $\bf {S}$ and ${\widetilde H}=\rho_{{\widetilde H}}({\bf S})$. In fact, assume that to strings $\Phi_{1}, \Phi_{2}$ and to a number ${\lambda}$ it corresponds respectively linear functionals ${\widetilde \varphi}_{\lambda 1}, {\widetilde \varphi}_{\lambda 2}$ on the eigenspace ${\bf S}^{\ast}_{\lambda}$ of $\bf {{\widehat A}}$ taking values $F_{\lambda}(\Phi_{1}), F_{\lambda}(\Phi_{2})$ on the element $F_{\lambda}$. Then to the string $\Phi=\alpha \Phi_{1}+\beta \Phi_{2}$ and to the same number $\lambda$ it corresponds the functional ${\widetilde \varphi}_{\lambda}$ on ${\bf S}^{\ast}_{\lambda}$ taking the value $F_{\lambda}(\alpha \Phi_{1}+\beta \Phi_{2})=\alpha F_{\lambda}(\Phi_{1})+\beta F_{\lambda}(\Phi_{2})$. That means that $\rho_{\widetilde H}$ is linear. The fact that $\rho_{\widetilde H}$ is injective follows from completeness of the set of eigenvectors $F_{\lambda}$ of ${\widehat {\bf A}}$. In fact, if $\rho_{\widetilde H}(\Phi)=0$, then ${\widetilde \varphi}_{\lambda}=0$, thus, $\Phi=0$.

The isomorphism $\rho_{\widetilde H}$ induces a Hilbert structure on ${\widetilde H}=\rho_{\widetilde H}({\bf S})$. Therefore it becomes an isomorphism of Hilbert spaces $\bf {S}$ and ${\widetilde H}$.

{\flushleft {\bf Theorem 6.8} {\em The string basis $e_{\widetilde H}=\rho^{-1}_{\widetilde H}$ is proper.}
\smallskip

{\it Proof}. To verify, notice that if
$\Phi \longrightarrow {\widetilde \varphi}_{\lambda}$ is the spectral decomposition of $\Phi$, then the spectral decomposition of $\Psi={\bf {\widehat A}}\Phi$ is ${\widetilde \psi}_{\lambda}=\lambda {\widetilde \varphi}_{\lambda}$. In fact, for any functional $F_{\lambda} \in {\bf S}^{\ast}_{\lambda}$ we have
\begin{equation}
F_{\lambda}(\Psi)=F_{\lambda}({\bf {\widehat A}}\Phi)=\lambda F_{\lambda}(\Phi),
\end{equation}
so that
\begin{equation}
{\widetilde \psi}_{\lambda}=\lambda {\widetilde \varphi}_{\lambda}.
\end{equation}
We therefore conclude that
\begin{equation}
F({\bf {\widehat A}} \Phi)=F(e^{-1}_{\widetilde H}e_{\widetilde H}{\bf {\widehat A}}\Phi)=
(e^{-1}_{\widetilde H})^{\ast}F(\lambda {\widetilde \varphi}_{\lambda})={\widetilde f}(\lambda {\widetilde \varphi}_{\lambda}),
\end{equation}
where ${\widetilde f}=(e^{-1}_{\widetilde H})^{\ast}F$. By (\ref{eb0}) this means that the basis $e_{\widetilde H}$ is proper.}
\bigskip

Moreover, in the considered case the complete set of eigenvectors $F_{\lambda}$ of ${\bf {\widehat A}}$ belongs to ${\bf S}^{\ast}$. Therefore, the complete set of eigenvectors of the operator of multiplication by $\lambda$ on ${\widetilde H}$ belongs to ${\widetilde H}^{\ast}$.

{\flushleft  {\bf Definition 6.9}. If a proper basis $e_{H}$  is such that the complete set of eigenvectors of the operator ${\widehat A}$ in this basis belongs to $H^{\ast}$ (alternatively, the complete set of eigenvectors of ${\bf {\widehat A}}$ belongs to ${\bf S}^{\ast}$), then the basis $e_{H}$ is called the {\it string basis of eigenvectors of ${\bf {\widehat A}}$} or the {\it string eigenbasis of ${\bf {\widehat A}}$}.}
\bigskip

We have thus proven the following theorem.

{\flushleft {\bf Theorem 6.10}} {\em If a complete set of eigenvectors $F_{\lambda}$ of an operator ${\bf {\widehat A}}$ is contained in ${\bf S}^{\ast}$, then there exists a string eigenbasis of ${\bf A}$.
That is, assume $F_{\lambda}({\bf {\widehat A}}\Phi)=\lambda F_{\lambda}(\Phi)$,
where the eigenvectors $F_{\lambda}$ of ${\bf {\widehat A}}$ form a complete set in ${\bf S}^{\ast}$. Then there exists a string basis $e_{\widetilde H}: {\widetilde H} \longrightarrow {\bf S}$ such that
for {\it any} functional $F \in {\bf S}^{\ast}$ and any string $\Phi$ we have
$F({\bf A}\Phi)={\widetilde f}(\lambda {\widetilde \varphi}_{\lambda})$,
where ${\widetilde \varphi}_{\lambda}$ and ${\widetilde f}$ are coordinates of $\Phi$ and $F$ in the basis $e_{\widetilde H}$ and its dual respectively.}
\bigskip

%%%%%%%%%%%%%%%%%%
%%%%%%%%%%%%%%%%%%%

It is well known that many operators useful in applications are not bounded. 
However, by an appropriate choice of metric on the image of a linear operator ${\widehat A}$ one can ensure the continuity of ${\widehat A}$. In other words, one can consider an unbounded operator ${\widehat A}$ on a Hilbert space $H$ as a bounded operator which maps $H$ into another Hilbert space ${\widetilde H}$. 
In particular, we have the following

{\flushleft {\bf Theorem 6.11} {\em Let ${\widehat A}$ be an invertible (possibly unbounded) operator on a dense subset $D({\widehat A}) \subset L_{2}$ having a range $R({\widehat A})\supset D({\widehat A})$. Then there exists a Hilbert metric on $R({\widehat A})$, in which ${\widehat A}$ is a bounded operator from $D({\widehat A})$ onto $R({\widehat A})$. }
\smallskip

{\it Proof}. For any $f, g \in R({\widehat A})$ define the inner product $(f,g)_{H}$ by
\begin{equation}
\label{A-metric}
(f,g)_{H}=({\widehat A}^{-1}f, {\widehat A}^{-1}g)_{L_{2}}.
\end{equation}
Then $\left\|{\widehat A}f\right\|_{H}=\left\|f\right\|_{L_{2}}$ for any $f \in D({\widehat A})$. That is, ${\widehat A}$ is bounded as a map from $D({\widehat A})$ with the metric $L_{2}$ onto $R({\widehat A})$ with the metric (\ref{A-metric}). Note that when ${\widehat A}$ is extended to the entire $L_{2}$ it becomes an isomorphism from $L_{2}$ onto the Hilbert space $H$ which is a completion of $R({\widehat A})$ in the metric (\ref{A-metric}).}
\bigskip

Let us also remark that if $e_{H}$ and $e_{{\widetilde H}}$ are string bases, the realization ${\widehat A}=e^{-1}_{{\widetilde H}}{\bf {\widehat A}}e_{H}$ of a continuous operator ${\bf {\widehat A}}: {\bf S} \longrightarrow {\bf S}$ may not be continuous when considered as a map into $H$. In other words, the unbounded operators in a Hilbert space $H$ may be realizations of continuous operators on ${\bf S}$.

Notice finally that when a realization ${\widehat A}$ of operator ${\bf {\widehat A}}:{\bf S} \longrightarrow {\bf S}$ is a map between two different Hilbert spaces, it becomes useful to generalize the {\em Definition 6.1} of the proper basis.
Namely, for any $F \in {\bf S}^{\ast}, \Phi \in {\bf S}$ and ${\bf {\widehat A}}:{\bf S}\longrightarrow {\bf S}$ we have
\begin{equation}
F({\bf {\widehat A}}\Phi)=F({\bf {\widehat A}}e_{H}\varphi)=F(e_{\widetilde {H}}e^{-1}_{\widetilde {H}}{\bf {\widehat A}}e_{H}\varphi)=e^{\ast}_{\widetilde {H}}F(e^{-1}_{\widetilde {H}} {\bf {\widehat A}} e_{H}\varphi)=f({\widehat A}\varphi),
\end{equation}
where $f=e^{\ast}_{\widetilde {H}}F \in {\widetilde H}^{\ast}$ and ${\widehat A}:H \longrightarrow {\widetilde H}$ is given by ${\widehat A}=e^{-1}_{\widetilde {H}} {\bf {\widehat A}} e_{H}$.
We then have the following
{\flushleft {\bf Definition 6.12} The realization ${\widehat A}=e^{-1}_{\widetilde {H}} {\bf {\widehat A}} e_{H}$ of ${\bf {\widehat A}}$ is called {\it proper} if for any $F \in {\bf S}^{\ast}$, $\Phi \in {\bf S}$ we have
\begin{equation}
\label{proper}
F({\bf {\widehat A}}\Phi)=f({\widehat A}\varphi)=f(\lambda \varphi),
\end{equation}
where $f=e^{\ast}_{\widetilde{H}}F$, $\varphi=e^{-1}_{H}\Phi$ and $\lambda$ is a function such that the operator of multiplication by $\lambda$ maps $H$ into $\widetilde{H}$.
}
\bigskip

{\flushleft If $H={\widetilde H}$ this is equivalent to the {\em Definition 6.1}. }

\section{Isomorphisms of Hilbert spaces preserving locality of operators}
\setcounter{equation}{0}
 
We saw that a single eigenvalue problem for an operator
on the string space leads to a family of eigenvalue problems in particular
string bases. Obviously, it is a general feature of tensor equations in the formalism:
a specific functional form of an equation depends on the choice of functional coordinates.

In the process of changing coordinates one may desire to preserve some specific properties of the equation. It becomes then important to describe all coordinate transformations preserving these properties. 

{\flushleft {\bf Definition 7.1} An operator ${\widehat A}$ on a Hilbert space $H$ is called {\em differential} or {\em local} if
\begin{equation}
\label{local}
{\widehat A}=\sum_{\left| q\right|\leq r}a_{q}D^{q}.
\end{equation}
Here $x \in R^{n}$,
$r$ is a nonnegative integer,
$q=(q_{1},...,q_{n})$ is a set of
nonnegative integers, $\left| q\right|=q_{1}+...+q_{n}$, the coefficients $a_{q}$ are functions of $x=(x_{1},...,x_{n})$ and
$D^{q}=\frac{\partial ^{|q|}}{\partial{x_{1}}^{q_{1}}...\partial{x_{n}} ^{q_{n}}}$.} 
\bigskip

It is easy to see that locality of an operator is not a coordinate invariant property. That is, the operator which is local in one system of functional coordinates does not have to be local in a different system of coordinates. At the same time, the property of being local is extremely important in applications. 

Let us describe the coordinate transformations $\omega:{\widetilde H}\longrightarrow H$ mapping differential operators on $H$ into differential operators on a Hilbert space ${\widetilde H}$.
We assume here that $H, {\widetilde H}$ are either spaces of sufficiently smooth functions of bounded support (or sufficiently fast decreasing at infinity), or dual to such spaces. 

All such transformations can be found by solving the equation
\begin{equation}
\omega^{-1}{\widehat A}\omega={\widehat B},
\end{equation}
where ${\widehat A}, {\widehat B}$ are appropriate differential operators. In expanded form we have:
\begin{equation}
\label{A}
\omega^{-1}\sum_{\left| q\right|\leq r}a_{q}D^{q}\omega=\sum_{\left| q\right|\leq s}b_{q}D^{q}.
\end{equation}
The latter equation can be also written in terms of the kernels of ${\widehat A}$ and ${\widehat B}$:
\begin{equation}
\label{local1}
\int \sum_{\left| q\right|\leq r}a_{q}(x)D^{q}\delta(z-x)\omega(z,y)dz=
\int \sum_{\left| q\right|\leq s}\omega(x,z)b_{q}(z)D^{q}\delta(y-z)dz.
\end{equation} 

{\flushleft {\bf Example}.} Let  $H, {\widetilde H}$ are spaces of functions of a single variable, ${\widehat A}=aD$,${\widehat B}=b$, where $a, b$ are functions, the equation (\ref{A}) reduces to 
\begin{equation}
\label{local10}
aD\omega=\omega b.
\end{equation}
By solving this differential equation we see that the kernel $\omega(x,y)$ of $\omega$ has the form
\begin{equation}
\label{solution10}
\omega(x,y)=g(y)e^{c(x)b(y)},
\end{equation}
where $c(x)=-\int\frac{dx}{a(x)}$ and $g$ is an arbitrary smooth function. To be a coordinate transformation $\omega$ must be an isomorphism as well.
In particular, the Fourier transform is a solution of (\ref{local10}) with
\begin{equation}
\omega(x,y)=e^{ixy}.
\end{equation}

More generally, if ${\widehat A}$, ${\widehat B}$ contain only one term each, the equation (\ref{local1}) yields
\begin{equation}
\label{local-gen}
a(x)\frac{\partial^{n} \omega(x,y)}{\partial x^{n}}=
(-1)^{m}\frac{\partial^{m} \left(\omega(x,y)b(y)\right)}{\partial y^{m}}.
\end{equation}

{\flushleft {\bf Example}.} Consider the equation (\ref{local-gen}) with $n=m=1$ assuming $a(x)$ and $b(y)$ are functions.
In this case the equation reads
\begin{equation}
\label{nonconst}
a(x)\frac{\partial{\omega(x,y)}}{\partial{x}}+\frac{\partial{(\omega(x,y)b(y))}}{\partial{y}}=0.
\end{equation}
Let us look for a solution in the form
\begin{equation}
\omega(x,y)=e^{f(x)g(y)}.
\end{equation}
Then (\ref{nonconst}) yields
\begin{equation}
\label{nonconst1}
a(x)f'(x)g(y)+b(y)f(x)g'(y)+b'(y)=0.
\end{equation}
If $b(y)=1$, (\ref{nonconst1}) is a separable equation and we have
\begin{equation}
\frac{a(x)f^{\prime }(x)}{f(x)}=-\frac{g^{\prime }(y)}{g(y)}=C,
\end{equation}
where $C$ is a constant. Solving this we obtain,
\begin{equation}
\omega(x,y)=e^{Ce^{\int \frac{C_{1}}{a(x)}dx} e^{-C_{1}y}}.
\end{equation}
Taking for example $C=C_{1}=1$ and $a(x)=x$, we have
\begin{equation}
\omega(x,y)=e^{xe^{-y}}.
\end{equation}
The corresponding transformation changes the operator $xD$ into the operator $D$.

\section{Isomorphisms of Hilbert spaces preserving the derivative operator}
\setcounter{equation}{0}

Among solutions of (\ref{local-gen}) those preserving the derivative operator $D$
are of particular interest. To find them consider the equation (\ref{local-gen}) with $n=m=1$ and with the constant coefficients $a=b=1$. Then (\ref{local-gen}) yields the following equation:
\begin{equation}
\label{derivative}
\frac{\partial \omega(x,y)}{\partial x}+\frac{\partial \omega(x,y)}{\partial y}=0.
\end{equation}
The smooth solutions of (\ref{derivative}) are given by
\begin{equation}
\label{der-sol}
\omega(x,y)=f(x-y),
\end{equation}
where $f$ is an arbitrary infinitely differentiable function on $R$.
In particular, the function
\begin{equation}
\label{exp}
\omega(x,y)=e^{-(x-y)^{2}}
\end{equation}
satisfies (\ref{derivative}). Also, we saw in section 2 that the corresponding
transformation considered on an appropriate Hilbert space ${\widetilde H}$ is injective and induces a Hilbert structure on the image $H$.
Therefore, it provides an example of a coordinate transformation that
preserves the derivative operator $D$ and, more generally, $D^{q}$.

When $H, {\widetilde H}$ are spaces of functions of $n$ variables, a similar role is played by the function
\begin{equation}
\label{exp1}
\omega(x,y)=e^{-(x-y)^{2}}
\end{equation}
with $x=(x_{1},...,x_{n})$, $y=(y_{1},...,y_{n})$  and the standard Euclidean metric on the space of variables. 

{\flushleft {\bf Theorem 8.1} {\em Let $L$ be a polynomial function of $n$ variables. Let $u, v \in \widetilde{H}$ be
functionals on the space $K$ of functions of $n$ variables which are infinitely differentiable
and have bounded supports.
Assume that $u$ is a generalized solution of
\begin{equation}
L\left( \frac{\partial}{\partial{x}_{1}},...,\frac{\partial}{\partial{x}_{n}}\right)u=v.
\end{equation}
Then there exists a smooth solution $\varphi$ of
\begin{equation}
L\left( \frac{\partial}{\partial{x}_{1}},...,\frac{\partial}{\partial{x}_{n}}\right)\varphi
=\psi,
\end{equation}
where $\varphi=\omega{u}$, $\psi=\omega{v}$ and $\omega$ is as in (\ref{exp1}).}
\smallskip

{\it Proof}.} Consider first the case of the ordinary differential equation
\begin{equation}
\label{diffeq}
\frac{d}{dx}u(x)=v(x).
\end{equation}
Assume $u$ is a generalized solution of (\ref{diffeq}). Define $\varphi=\omega{u}$ and
$\psi=\omega{v}$, where $\omega$ is as in (\ref{exp}). Notice that $\varphi, \psi$ are
infinitely differentiable. In fact, any functional on the space $K$ of infinitely
differentiable functions of bounded support acts as follows (see \cite{Gelfand1}):
\begin{equation}
(f,\varphi )=\int F(x)\varphi ^{(m)}(x)dx,
\end{equation}
where $F$ is a continuous function on $R$. Applying $\omega$ to $f$ shows that the result
is a smooth function.

As $\omega^{-1}\frac{d}{dx}\omega=\frac{d}{dx}$, we have
\begin{equation}
\label{diffgen}
\omega^{-1}\frac{d}{dx}\omega{u}=v.
\end{equation}
That is,
\begin{equation}
\frac{d}{dx}\varphi(x)=\psi(x)
\end{equation}
proving the theorem in this case.
The higher order derivatives can be treated similarly as
\begin{equation}
\label{higherder}
\omega^{-1}\frac{d^n}{dx^{n}}\omega=
\omega^{-1}\frac{d}{dx}\omega \omega^{-1}\frac{d}{dx}\omega...
\omega^{-1}\frac{d}{dx}\omega.
\end{equation}
That is, transformation $\omega$ preserves derivatives of any order.
Generalization to the case of several variables is straightforward.

\section {Transformations preserving a product of functions}
\setcounter{equation}{0}

Consider a simple algebraic equation
\begin{equation}
\label{product0}
a(x)f(x)=h(x),
\end{equation}
where $f$ is an unknown (generalized) function of a single variable $x$ and $h$ is an element of a Hilbert space $H$ of functions on a set $D \subset R$.
To investigate transformation properties of this equation we need to interpret
it as a tensor equation on the string space $\bf {S}$. The right hand side is
a function. Therefore this must be a ``vector equation" (i.e. both sides must be $(1,0)$-tensors
on the string space).
If $f$ is to be a function as well, $a$ must  be a $(1,1)$-tensor. 

To preserve the product-like form of the equation we need such a coordinate transformation
$\omega:\widetilde{H}\longrightarrow H$ that
\begin{equation}
\label{product}
\omega^{-1}a\omega=b.
\end{equation}
In this case the equation (\ref{product0}) in new coordinates is
\begin{equation}
b(x)\varphi(x)=\psi(x),
\end{equation}
where $h=\omega{\psi}$, $f=\omega{\varphi}$,
and $\varphi,\psi \in \widetilde{H}$.

Equation (\ref{product}) is clearly satisfied whenever $a=b=C$, where $C$ is a constant function. On another hand, whenever $b' \neq 0$ and $H$ contains sufficiently many functions, we deduce as in the {\em Theorem 6.5} that
$\omega(x,y)=d(x)\delta(x-y)$ for some function $d$. 

The obtained result then says that the operator of multiplication can be preserved
only in trivial cases when $a(x)=C$ or $\omega$ itself is an operator of
multiplication by a function. 

In particular, the product of non-constant functions of one and the same variable is not
an invariant operation under a general transformation of functional coordinates.

\section{Coordinate transformations of nonlinear functional tensor equations}
\setcounter{equation}{0}

It is known that the theory of generalized functions has been mainly successful with
linear problems. The difficulty of course lies in defining the product of generalized functions. We saw in the previous section that multiplication of functions of the same variable is not an invariant operation. The idea is then to define an invariant operation which reduces to multiplication in particular coordinates. 

More generally, the developed functional coordinate formalism offers a systematic approach for dealing with nonlinear equations in generalized functions. Namely, the terms in a nonlinear functional tensor equation  represent tensors. Because of that the change of coordinates is meaningful and can be used to extend nonlinear operations to generalized functions. 
{\flushleft {\bf Example}.} Consider a nonlinear equation with the term $(\varphi(x))^{2}$.
To interpret this term as a functional tensor we write
\begin{equation}
\label{nonlin1}
\varphi(x)\cdot \varphi(x)=
\int \delta(x-u)\delta(x-v)\varphi(u)\varphi(v)dudv.
\end{equation}
Therefore, this term can be considered to be the convolution of the $(1,2)$-tensor
\begin{equation}
\label{nonlin2}
c^{x}_{uv}=\delta(x-u)\delta(x-v)
\end{equation}
with the pair of strings $\varphi^{u}=\varphi(u)$:
\begin{equation}
\varphi(x)\cdot \varphi(x)=c^{x}_{uv}\varphi^{u}\varphi^{v}.
\end{equation}
With this identification we can now apply a coordinate transformation to make $\varphi$ singular at the expense of smoothing $c^{x}_{uv}$. In particular, we can transform $\varphi$ into the delta-function.

{\flushleft \bf{Example.}} Consider the equation
\begin{equation}
\label{nonlin7}
\int k(x-y)\frac {d\varphi_{t}(x)}{d t}\frac {d\varphi_{t}(y)}{d t}dxdy=0,
\end{equation}
where $\varphi_{t}(x)$ is an unknown function which depends on the parameter $t$ and $k(x,y)$ is a smooth function on $R^{2n}$.
   
Let us look for a solution in the form $\varphi_{t}(x)=\delta(x-a(t))$. As
\begin{equation}
\label{1A}
\frac {d\varphi_{t}(x)}{d t}=-\frac {\partial \delta(x-a(t))}{\partial x^{\mu}}\frac {d a^{\mu}}{d t},
\end{equation}
we have after ``integration by parts" the following equation:
\begin{equation}
\label{nonlin8}
\left .\frac {\partial^{2}k(x,y)}{\partial x^{\mu} \partial y^{\nu}}\right |_{x=y=a(t)} \frac {d a^{\mu}(t)}{d t}\frac {d a^{\nu}(t)}{d t}=0.
\end{equation}
We remark here that as explained in section 11, the formula (\ref{1A}) and the method of ``integration by parts'' in (\ref{nonlin7}) are valid.
If the tensor field 
\begin{equation}
g_{\mu \nu}(a)\equiv \left .\frac {\partial^{2}k(x,y)}{\partial x^{\mu} \partial y^{\nu}}\right |_{x=y=a}
\end{equation}
 is symmetric and positive definite, the equation (\ref{nonlin8}) has only the trivial solution. However, if $g_{\mu \nu}(a)$ is non-degenerate and indefinite, then there is a nontrivial solution. In particular, we can choose $g_{\mu \nu}$ to be the Minkowski tensor $\eta_{\mu\nu}$ on space-time. For this assume that $x,y$ are space-time points and take 
\begin{equation}
k(x,y)=e^{-\frac{(x-y)^2}{2}},
\end{equation}
where $(x-y)^{2}=\eta_{\mu\nu}(x-y)^{\mu}(x-y)^{\nu}$. Then we immediately conclude that $g_{\mu\nu}$ is the Minkowski tensor $\eta_{\mu\nu}$.
Solutions to (\ref{nonlin8}) are then given by the null lines $a(t)$. Therefore, the original problem (\ref{nonlin7}) has solutions of the form $\varphi_{t}(x)=\delta(x-a(t))$ where $a(t)$ is a null line.

Notice, that the obtained generalized functions $\varphi_{t}$ are singular generalized solutions to the nonlinear equation (\ref{nonlin7}). Once again, these solutions become possible because the kernel $k(x,y)$ in (\ref{nonlin7}) is a smooth function, so the convolution $k_{xy}\dot {\varphi}^{x}\dot {\varphi}^{y}$ is meaningful.

{\flushleft \bf{Example.}} Consider the equation
\begin{equation}
\int e^{-\frac{(x-y)^2}{2}} \quad \frac {d\varphi_{t}(x)}{d t}\frac {d\varphi_{t}(y)}{d t}dxdy=1,
\end{equation}
where the metric on the space of variables is Euclidean: $(x-y)^{2}=\delta_{\mu\nu}(x-y)^{\mu}(x-y)^{\nu}$. Looking for solution in the form $\varphi_{t}(x)=\delta(x-a(t))$ and performing ``integration by parts'', we obtain
\begin{equation}
\label{nonlin8A}
\delta_{\mu \nu}\frac {d a^{\mu}(t)}{d t}\frac {d a^{\nu}(t)}{d t}=1.
\end{equation}
That is, $\left \|\frac{d a(t)}{dt}\right\|_{R^{n}}=1$. Therefore, $\varphi_{t}(x)=\delta(x-a(t))$, where the path $a(t)$ has a unit velocity vector at any $t$.

Let us apply a transformation $\rho$ with kernel $\rho(x,y)=e^{-(x-y)^{2}}$ to the equation
\begin{equation}
\int e^{-\frac{(x-y)^2}{2}} \quad \frac {d \delta(x-a(t))}{d t}\frac {d\delta(y-a(t))}{d t}dxdy=1.
\end{equation}
This yields
\begin{equation}
M\int \delta(x-y) \frac {d e^{-(x-a(t))^{2}}}{d t}\frac {d e^{-(y-a(t))^{2}}}{d t}dxdy=1,
\end{equation}
where $M$ is a constant. This is the same equation in the coordinates in which the kernel of the metric is the delta function, while the solution given originally by the delta-function becomes a smooth exponential function. 

\section{Embeddings of $n$-dimensional manifolds into Hilbert spaces of functions }
\setcounter{equation}{0}

In this section we will use various dual Hilbert spaces $H, H^{\ast}$, where the elements of $H^{\ast}$ are smooth functions and the metric on the dual space $H$ is given by a smooth on $R^{n}\times R^{n}$ kernel $k(x,y)$. Examples of such spaces were given in section 2.

{\flushleft {\bf Theorem 11.1} {\em The space $H$ described above contains the evaluation functionals $\delta(x-a)$ for all $a$ in $R^{n}$. }
\smallskip

{\it Proof}.} Because $k(x,y)$ is smooth, the sequence $f_{L}=\left (\frac{L}{\sqrt {\pi}}\right )^{n}e^{-L^{2}(x-a)^{2}}$, $L=1,2,3,...$ is fundamental in $H$. As $H$ is complete, $f_{L}$ converges to an element $f$ in $H$. Therefore, $f_{L}$ converges to $f$ weakly in $H$ as well. By Riesz theorem this is equivalent to saying that $(f_{L}, \varphi)\longrightarrow (f,\varphi)$ for all $\varphi \in H^{\ast}$. At the same time, $f_{L}$ is a delta-converging sequence. In particular, $(f_{L}, \varphi)\longrightarrow (\delta_{a},\varphi)$ for all $\varphi \in H^{\ast}$. The uniqueness of the weak limit signifies then that $f_{L}$ converges to $\delta_{a}$ in $H$.  
\bigskip  

{\flushleft { {\bf Theorem 11.2}} {\em If the space $H^{\ast}$ contains sufficiently many functions,  the subset $M$ of all delta-functions in $H$ forms a $n$-dimensional submanifold of $H$ diffeomorphic to $R^{n}$. }
\smallskip

{\it Proof}.} The map $\omega: a \longrightarrow \delta(x-a)$ is a smooth from $R^{n}$ into $H$. Assume that for any two points $a, b \in R^{n}$ there is a function $\varphi \in H^{\ast}$ such that $\delta_{a}(\varphi) \neq \delta_{b}(\varphi)$. This is true in particular if $H^{\ast}$ contains all $C^{\infty}$-functions of bounded support. In  this case the map $\omega$ is injective. The smooth injective map $\omega$ parametrizes the set $M$ of all delta-functions in $H$ identifying $M$ with a submanifold of $H$ diffeomorphic to $R^{n}$.
\bigskip

Note that, although $M$ is not a linear subspace of $H$, the diffeomorphism $\omega: R^{n} \longrightarrow M$ induces a linear structure on $M$. In fact, we can define linear operations $\oplus,\odot$ on $M$ by $\omega(x+y)=\omega(x) \oplus \omega(y)$ and $\omega (kx)=k \odot \omega (x)$ for any vectors $x,y \in R^{n}$ and any number $k$. 

{\flushleft  {\bf Theorem 11.3} {\em These operations are continuous in topology of $H$. }
\smallskip

{\it Proof}}. $\int k(x,y) (\lambda \delta(x-a)-\lambda_{k}\delta(x-a_{k}) (\lambda \delta(y-a)-\lambda_{k}\delta(y-a_{k})dxdy=\lambda^{2}k(a,a)-\lambda \lambda_{k}(k(a,a_{k})+k(a_{k},a))+\lambda^{2}_{k}k(a_{k},a_{k}) \longrightarrow 0$ provided $a_{k}\longrightarrow a$ and $\lambda_{k} \longrightarrow \lambda$. The continuity of addition is verified in a similar way. 
\bigskip

By using the same method one can also derive topologically nontrivial spaces $M$. For example,
let $H^{\ast}$ be the Hilbert space of smooth functions on the interval $[0,2\pi]$ which contains all smooth functions of bounded support in $(0, 2\pi)$. Assume further that 
\begin{equation}
\label{phi-pi}
\varphi^{(n)}(0)=\varphi^{(n)}(2\pi)
\end{equation} 
for all $\varphi$ in $H^{\ast}$ and for all orders $n$ of (one-sided) derivatives of $\varphi$. Consider the dual space $H$ of functionals in $H^{\ast}$ and assume that the kernel of the metric on $H$ is a smooth function.

{\flushleft { {\bf Theorem 11.4}} The subset $M$ of delta-functions in $H$ form a submanifold diffeomorphic to the circle $S^{1}$.}
\smallskip

{\it Proof}. The map $\omega: a \longrightarrow \delta(\theta -a)$ from $(0, 2\pi)$ into $H$ is $C^{\infty}$. It is also injective, because $H^{\ast}$ contains sufficiently many functions to distinguish any two delta-functions $\delta(x-a)$, $\delta(x-b)$ with $a,b \in (0,2\pi)$, $a\neq b$. Finally, the functionals $\delta(\theta), \delta(\theta-2\pi)$ are identical on $H^{\ast}$ as $\int \delta(\theta-2\pi)\varphi(\theta)d\theta=\varphi(2\pi)=\varphi(0)=\int \delta(\theta)\varphi(\theta)d\theta$ for all $\varphi$ in $H^{\ast}$.
\bigskip

More generally, a Hilbert space $H^{\ast}$ of functions on an n-dimensional manifold can be identified with the space of functions on a subset of $R^{n}$. In fact, the manifold itself is a collection of non-intersecting ``pieces'' of $R^{n}$ ``glued'' together. Functions on the manifold can be then identified with functions defined on the disjoint union of all pieces and taking equal values at the points identified under ``gluing''. As a result, the dual space $H$ of generalized functions ``on'' the manifold can be also identified with the corresponding space of generalized functions ``on'' a subset of $R^{n}$.

This fact allows us to conclude that topologically different manifolds $M$ can be obtained by choosing an appropriate Hilbert space $H$ of functions ``on'' a subset of $R^{n}$ and identifying $M$ with the submanifold of $H$ consisting of delta-functions. The manifold structure on $M$ is then induced by the inclusion of $M$ into $H$ and does not have to be defined in advance. In particular, the ``gluing conditions'' like (\ref{phi-pi}) imposed on functions result in the corresponding ``gluing'' of the delta-functions and of the appropriate subsets of $R^{n}$.

Moreover, the tangent bundle structure and the Riemannian structure on $M$ can be also induced by the embedding $i: M \longrightarrow H$. This embedding is {\em natural} in a sense that the formalism of local coordinates on $M$ turns out to be a ``restriction'' to $M$ of the developed functional coordinate formalism.

In fact, given vector $X$ tangent to a path $\Phi_{t}$ in $\bf {S}$ at the point $\Phi_{0}$ and a differentiable functional $F$ on a neighborhood of $\Phi_{0}$ in $\bf {S}$, the directional derivative of $F$ at $\Phi_{0}$ along $X$ is defined by
\begin{equation}
\label{tangent}
XF=\left.\frac{dF(\Phi_{t})}{dt}\right|_{t=0}.
\end{equation}
By applying the chain rule we have
\begin{equation}
\label{chain}
XF=
\left.F^{\prime}(\Phi)\right|_{\Phi=\Phi_{0}}\left.\Phi^{\prime}_{t}\right|_{t=0},
\end{equation}
where $F^{\prime}(\Phi)|_{\Phi=\Phi_{0}}: {\bf S} \longrightarrow R$ is the derivative functional at $\Phi=\Phi_{0}$ and $\Phi^{\prime}_{t}|_{t=0} \in {\bf S}$ is the derivative of $\Phi_{t}$ at $t=0$. Let $e_{H}$ be a functional basis on ${\bf S}$. Writing (\ref{chain}) in coordinates $({\bf S}, e^{-1}_{H})$ yields
\begin{equation}
\label{var-d}
XF=
\int \left. \frac{\delta f(\varphi)}{\delta \varphi(x) }\right|_{\varphi=\varphi_{0}}\xi (x)dx,
\end{equation}
where $\xi=\varphi^{\prime}_{t}|_{t=0}$, $\varphi_{t}=e^{-1}_{H}\Phi_{t}$, and the linear functional
$\left.\frac{\delta f(\varphi)}{\delta \varphi (x) }\right|_{\varphi=\varphi_{0}} \in H^{\ast}$, is the derivative functional $F^{\prime}(\Phi_{0})$ in the dual basis $e^{\ast}_{H}$.

Let us select from all paths in $H$ the paths with values in $M$. In the chosen coordinates any such path $\varphi_{t}: [a,b] \longrightarrow M$  has the form
\begin{equation}
\label{path1}
\varphi_{t}(x)=\delta(x-a(t))
\end{equation}
for some function $a(t)$ taking values in $R^{n}$.

%%%%%%%%%%%%%%%%%%%%%%%%%%%%%%%%%%%%%%%%
{\flushleft {\bf Theorem 11.5} {\em The expression (\ref{tangent}) evaluated on appropriate functionals and on a path (\ref{path1}) yields the standard expression $\left.\frac{\partial f(a)}{\partial a^{\mu}}\right|_{a=a(0)}\left.\frac{da^{\mu}}{dt}\right|_{t=0}$ for action of the vector tangent to the path $a(t):[a,b] \longrightarrow R^{n}$ at $t=0$ on differentiable functions $f$.}
\smallskip

{\it Proof.}}   
Assume that $f$ is an analytic functional represented on a neighborhood of $\varphi_{0}=\left .\varphi_{t}\right |_{t=0}=\delta_{a(0)}$ in $H$
%on a ball in a ball
%$\left\|\varphi-\varphi_{0}\right\|_{H} \le \delta < \epsilon$ centered at $\varphi_{0}$ the functional $f$ can be represented 
by a convergent in $H$ power series 
\begin{equation}
f(\varphi)=f_{0}+\int f_{1}(x)\varphi(x)dx+\int \int f_{2}(x,y)\varphi(x)\varphi(y)dxdy+...\quad .
\end{equation}
Assume further that the function 
\begin{equation}
f(a)=f(\delta(x-a))=f_{0}+f_{1}(a)+f_{2}(a,a)+...\quad 
\end{equation}
is smooth on a neighborhood of $a_{0}=a(0)$ (for example, $f_{1}$ is a smooth function and $f_{k}=0$ for $k\ge 2$).
Then on the path $\varphi_{t}=\delta(x-a(t))$ we have
\begin{equation}
\label{4_vector}
\left.\frac{df(\varphi_{t})}{dt}\right|_{t=0}=
\left.\frac{\partial f(a)}{\partial a^{\mu}}\right|_{a=a(0)}\left.\frac{da^{\mu}}{dt}\right|_{t=0}.
\end{equation}
In particular, the expression on the right of (\ref{4_vector}) is the action of the n-vector $\frac{da^{\mu}}{dt} \frac{\partial}{\partial a^{\mu}}$ tangent to the path $a(t)$ on the smooth function $f(a)$. 
\bigskip

Assume once again that $H$ is a real Hilbert space and let $K: H \times H \longrightarrow R$ be the metric on $H$ given by a smooth kernel $k(x,y)$. 
If $\varphi=\varphi_{t}(x)=\delta(x-a(t))$ is a path in $M$, then for the vector $\delta \varphi(x)$ tangent to the path at $\varphi_{0}$ we have 
\begin{equation}
\delta \varphi(x)\equiv \left.\frac{d\varphi_{t}(x)}{dt}\right|_{t=0}=-\nabla_{\mu}\delta(x-a)\left.\frac {da^{\mu}}{dt}\right|_{t=0}. 
\end{equation}
Here $a=a(0)$ and derivatives are understood in a generalized sense.  Therefore,
\begin{equation}
\left \| \delta \varphi \right \|^{2}_{H}=
\int k(x,y) \nabla_{\mu}\delta(x-a)\left.\frac {da^{\mu}}{dt}\right|_{t=0}\nabla_{\nu}\delta(y-a)
\left.\frac {da^{\nu}}{dt}\right|_{t=0}dxdy.
\end{equation}
``Integration by parts" in the last expression gives
\begin{equation}
\label{fin}
\int k(x,y) \delta \varphi(x)\delta \varphi(y)dxdy=
\left.\frac {\partial^{2}k(x,y)}{\partial x^{\mu} \partial y^{\nu}}\right|_{x=y=a} \left.\frac {da^{\mu}}{dt}\right|_{t=0}\left.\frac {da^{\nu}}{dt}\right|_{t=0}.
\end{equation}

\bigskip

{\flushleft {\bf Remark}}. Although the above manipulations with generalized functions are somewhat formal, they can be easily justified. In particular, from {\em Theorem 11.1} we know that as $L\longrightarrow \infty$, the sequence $f_{L}(x-a)=\left( \frac{L}{\sqrt{\pi}}\right)^{n}e^{-L^{2}(x-a)^{2}}$ converges in norm in $H$ to $\delta(x-a)$. Similarly, the sequence of the derivatives $\nabla_{\mu} f_{L}(x-a)$ converges to $\nabla_{\mu}\delta(x-a)$. Performing now the ordinary integration by parts in 
\begin{equation}
\int k(x,y) \nabla_{\mu}f_{L}(x-a)\left.\frac {da^{\mu}}{dt}\right|_{t=0}\nabla_{\nu}f_{L}(y-a)
\left.\frac {da^{\nu}}{dt}\right|_{t=0}dxdy
\end{equation}
and taking $L$ to infinity we obtain the result (\ref{fin}).

\bigskip

By defining $\frac {da^{\mu}}{dt}|_{t=0}=da^{\mu}$, we now have
\begin{equation}
\label{relation1}
\int k(x,y) \delta \varphi(x)\delta \varphi(y)dxdy=g_{\mu \nu}(a)da^{\mu}da^{\nu},
\end{equation}
where
\begin{equation}
\label{metric}
g_{\mu \nu}(a)=\left.\frac {\partial^{2}k(x,y)}{\partial x^{\mu} \partial y^{\nu}}\right|_{x=y=a}.
\end{equation}

As the functional $K$ is symmetric, the tensor $g_{\mu \nu}(a)$ can be assumed to be symmetric as well. If in addition $\left.\frac{\partial^{2}k(x,y)}{\partial x^{\mu} \partial y^{\nu}}\right|_{x=y=a}$ is positive definite at every $a$, the tensor $g_{\mu \nu}(a)$ can be identified with the Riemannian metric on an $n$-dimensional manifold $M$ given in local coordinates $a^{\mu}$. 

%\newpage

{\flushleft {\bf Example.}} Consider the Hilbert space $H$ with metric given by the kernel $k({\bf x},{\bf y})=e^{-\frac{1}{2}({\bf x}-{\bf y})^{2}}$ for all ${\bf x},{\bf y} \in R^{3}$. Using (\ref{metric}) and assuming $({\bf x}-{\bf y})^{2}=\delta_{\mu\nu}(x^{\mu}-y^{\mu})(x^{\nu}-y^{\nu})$ with $\mu,\nu =1,2,3$, we immediately conclude that $g_{\mu\nu}(a)=\delta_{\mu\nu}$.

The resulting isometric embedding of the Euclidean space $N=R^{3}$ into $H$ is illustrated in Figure 1. The cones in the figure represent delta-functions forming the manifold $M$ which we denote in this case by $M_{3}$.

\begin{figure}[ht]
\label{fig:1}
\begin{center}
\includegraphics[width=8cm]{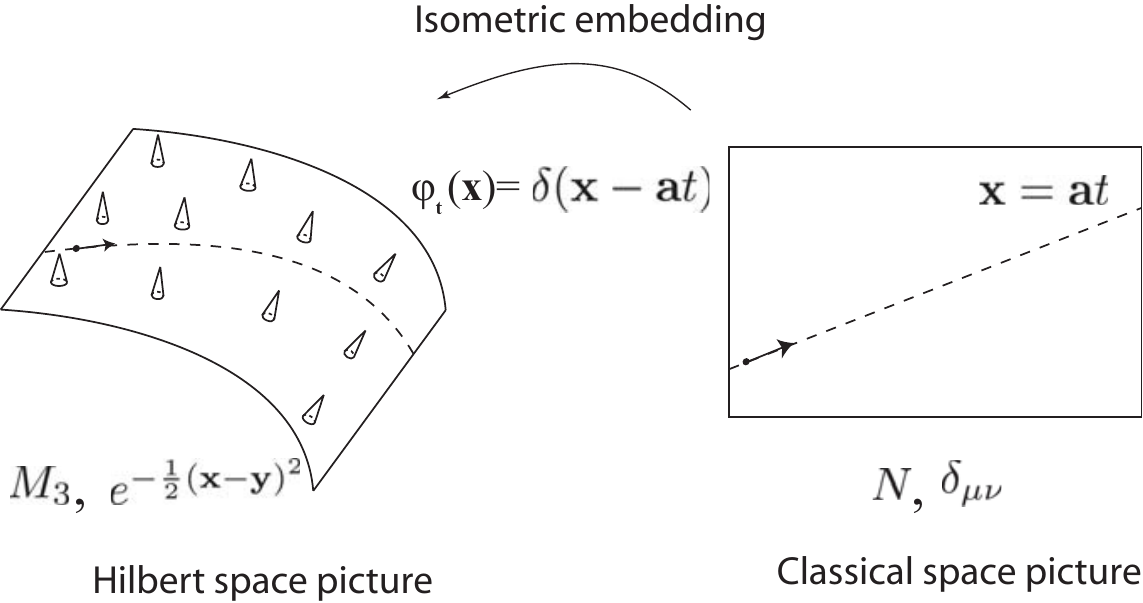}
\caption{\small{Isometric embedding of $R^{3}$ into $H$}}
\end{center}
\end{figure}

The following theorem clarifies the embedding of $M_{3}$ into the space $H$ with the metric given by $e^{-\frac{1}{2}(x-y)^{2}}$.
 
{\flushleft { {\bf Theorem 11.6}} {\em The manifold $M_{3}$ is a submanifold of the unit sphere $S^{H}$ in $H$. Moreover, the set $M_{3}$ form a {\em complete system} in $H$. The elements of any finite subset of $M_{3}$ are linearly independent.}
\smallskip

{\it Proof}.}
Observe that the norm of any element $\delta({\bf x}-{\bf a})$ in $H$ is equal to $1$. Therefore, the three dimensional manifold $M_{3}$ is a submanifold of the unit sphere $S^{H}$ in $H$.
 
To show that the set $M_{3}$ form a {\em complete system} in $H$ we need to verify that there is no non-trivial element of $H$ orthogonal to every element of $M_{3}$. Assume that $f$ is a functional in $H$ such that $\int e^{-\frac{1}{2}({\bf x}-{\bf y})^{2}}f({\bf x})\delta({\bf y}-{\bf u})d{\bf x}d{\bf y}=0$ for all ${\bf u} \in R^{3}$. Then $\int e^{-\frac{1}{2}({\bf x}-{\bf u})^{2}}f({\bf x})d{\bf x}=0$ for all ${\bf u} \in R^{3}$. 
Since the metric ${\widehat G}: H \longrightarrow H^{\ast}$ given by the kernel $e^{-\frac{1}{2}({\bf x}-{\bf y})^{2}}$ is an isomorphism, we conclude that $f=0$.

Assume now that  $\sum_{k=1}^{n} c_{k}\delta({\bf x}-{\bf a}_{k})$ is the zero functional in $H^{\ast}$ and the vectors ${\bf a}_{k} \in R^{3}$ are all different.
For any such finite set of vectors, the space $H^{\ast}$ contains functions $\varphi_{k}$, $k=1,...,n$ with supports containing one and only one of the points ${\bf a}_{k}$ each and so that no two supports contain the same point. 
Therefore the coefficients $c_{k}$ must be all equal to zero, that is, the elements of any finite subset of $M_{3}$ are linearly independent.
\bigskip 

Note that the set $M_{3}$ is uncountable and that no two elements of $M_{3}$ are orthogonal (although, provided $|{\bf a}-{\bf b}| \gg 1$, the elements $\delta({\bf x}-{\bf a})$, $\delta({\bf x}-{\bf b})$ are ``almost'' orthogonal).   

The following two pictures help ``visualizing'' the embedding of $M_{3}$ into $H$. Under the embedding any straight line ${\bf x}={\bf a_{0}}+{\bf a}t$ in $R^{3}$ becomes a ``spiral'' $\varphi_{t}({\bf x})=\delta({\bf x}-{\bf a_{0}}-{\bf a}t)$ on the sphere $S^{H}$ through dimensions of $H$. One such spiral is shown in Figure 2. The curve in Figure 2 goes through the tips of three shown linearly independent unit vectors. Imagine that each point on the curve is the tip of a unit vector and that any $n$ of these vectors are linearly independent. 

\begin{figure}[ht]
\label{fig:2}
\begin{center}
\includegraphics[width=2cm]{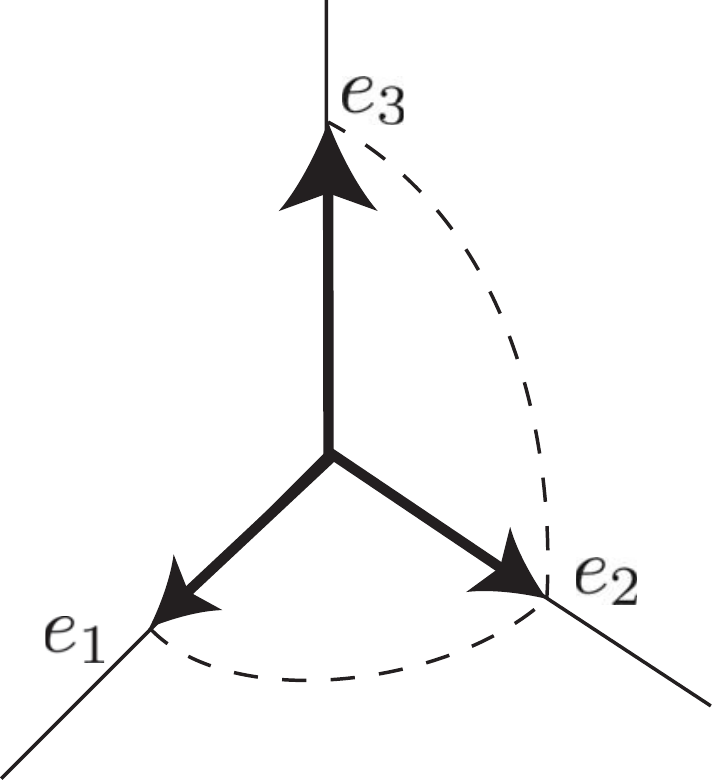}
\caption{\small{Straight line in $R^{3}$ as a ``spiral'' through dimensions of $H$}}
\end{center}
\end{figure}

The manifold $M_{3}$ is spanned by such spirals.
 Figure 3 illustrates the embedding of $M_{3}$ into $H$ in light of this result.

\begin{figure}[ht]
\label{fig:3}
\begin{center}
\includegraphics[width=5.5cm]{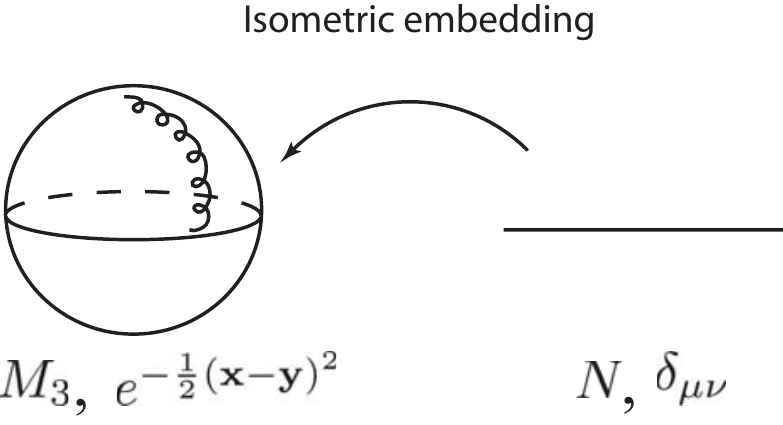}
\caption{\small{Embedding of $R^{3}$ into a unit sphere in $H$}}
\end{center}
\end{figure}

\section{Riemannian metric on the unit sphere in $L_{2}$ and on the complex projective space $CP^{L_{2}}$}

\setcounter{equation}{0}

In this section we apply the developed coordinate formalism to demonstrate that solutions of the Schr{\"o}dinger equation $\frac{d \varphi_{t}}{dt}=-i{\widehat h}\varphi_{t}$ for a closed quantum system described by the Hamiltonian ${\widehat h}$ are geodesics in the appropriate Riemannian metric on the space of states of the system. 
  
For this it will turn out to be convenient to use the index notation introduced in section 4. Thus, a string-tensor $T$ or rank $(r,s)$ in the index notation will be written as $t^{a_{1}...a_{r}}_{b_{1}...b_{s}}$. 
Assume that ${\widehat K}: H \longrightarrow H^{\ast}$ defines an Hermitian inner product $K(\xi,\eta)=(\widehat{K}\xi,\eta)$ on a complex Hilbert space $H$ of compex-valued functions $\xi$. Let $H_{R}$ be the real Hilbert space which is the {\it realization} of $H$. Namely, $H_{R}$ is the space of pairs of vectors $X=(\xi, {\overline \xi})$ with multiplication by real numbers. 

Since the inner product on $H$ is Hermitian, it defines a real valued Hilbert metric on $H_{R}$ by
\begin{equation}
K_{R}(X,Y)=2ReK(\xi,\eta),
\end{equation}
for all $X=(\xi, {\overline \xi})$, $Y=(\eta, {\overline \eta})$ with $\xi, \eta \in H$.
We will also use the ``matrix" representation of the corresponding operator ${\widehat K}_{R}: H_{R} \longrightarrow H_{R}^{\ast}$:
\begin{equation}
\label{metricKR}
\widehat {K}_{R}=\left[ 
\begin{array}{cc}
0 & {\widehat K} \\ 
{\overline {\widehat K}} & 0
\end{array}
\right].
\end{equation}
In particular, we have
\begin{equation}
K_{R}(X,Y)=({\widehat K}_{R}X,Y)=[\xi,{\overline \xi}]
{\widehat K}_{R}
\left[ 
\begin{array}{c}
\eta \\ 
{\overline \eta}
\end{array}
\right]=2Re({\widehat K}\xi,\eta),
\end{equation}
where $\xi{\widehat K}{\overline \eta}$ stands for the inner product $({\widehat K}\xi,\eta)$ and ${\overline \xi}{\overline{\widehat K}} \eta$ stands for its conjugate.

Let us agree to use the capital Latin letters $A, B, C, ...$ as indices of tensors defined on direct products of copies of the real Hilbert space $H_{R}$ and its dual. The small Latin letters $a, b, c, ...$ and the corresponding overlined letters ${\overline a}, {\overline b}, {\overline c}, ... $ will be reserved for tensors defined on direct products of copies of the complex Hilbert space $H$, its conjugate, dual and dual conjugate. A single capital Latin index replaces a pair of lower Latin indices. For example, if $X \in H_{R}$, then $X^{A}=(X^{a}, X^{{\overline a}})$, with $X^{a}$ representing an element of $H$ and  $X^{{\overline a}}={\overline X^{a}}$.  

Consider now the tangent bundle over a complex string space $\bf {S}$ which we identify here with a Hilbert space $L_{2}$ of square-integrable functions.  
Let us identify all fibers of the tangent bundle over $L_{2}$ (i.e. all tangent spaces $T_{\varphi}L_{2}$, $\varphi \in L_{2}$) with the complex Hilbert space $H$ described above. 
Let us introduce an Hermitian $(0,2)$ tensor field $G$ on the space $L_{2}$ without the origin as follows:
\begin{equation}
\label{Riem}
G(\xi,\eta)=\frac{({\widehat K}\xi,\eta)}{(\varphi,\varphi)_{L_{2}}},
\end{equation}
for all $\xi$, $\eta$ in the tangent space $T_{\varphi}L_{2}$ and all points $\varphi \in L_{2\ast}$. Here $L_{2\ast}$ stands for the space $L_{2}$ without the origin. 

The corresponding (strong) Riemannian metric $G_{R}$ on $L_{2}$ is defined by
\begin{equation}
G_{R}(X,Y)=2ReG(\xi,\eta),
\end{equation} 
where as before $X=(\xi, {\overline \xi})$ and $Y=(\eta, {\overline \eta})$.
In the matrix notation of (\ref{metricKR}) we have for the operator ${\widehat G}_{R}: H_{R} \longrightarrow H_{R}^{\ast}$ defining the metric $G_{R}$:
\begin{equation}
\label{metricGR_1}
{\widehat G}_{R}=\left[ 
\begin{array}{cc}
0 & {\widehat G} \\ 
{\overline {\widehat G}} & 0
\end{array}
\right],
\end{equation}
where ${\widehat G}: H \longrightarrow H^{\ast}$ defines the metric $G$.

In our index notation the kernel of the operator $\widehat {G}$ will be denoted by $g_{a{\overline b}}$, so that
\begin{equation}
\label{index_metric}
g_{a{\overline b}}=\frac{k_{a{\overline b}}}{\left\|\varphi \right\|^{2}_{L_{2}}},
\end{equation}
where $k_{a {\overline b}}$ is the kernel of ${\widehat K}$. 
From (\ref{metricGR_1}) we have for the components $({\widehat G}_{R})_{AB}$ of the metric ${\widehat G}_{R}$:
\begin{equation}
({\widehat G}_{R})_{ab}=({\widehat G}_{R})_{{\overline a}{\overline b}}=0,
\end{equation}
and
\begin{equation}
({\widehat G}_{R})_{a{\overline b}}=g_{a{\overline b}}, \quad
({\widehat G}_{R})_{{\overline a}b}={\overline g}_{a{\overline b}}.
\end{equation}
For this reason and with the agreement that $g_{{\overline a}b}$ stands for ${\overline g}_{a{\overline b}}$ we can denote the kernel of ${\widehat G}_{R}$ by $g_{AB}$. For the inverse metric we have
\begin{equation}
\label{metricGR}
{\widehat G}^{-1}_{R}=\left[ 
\begin{array}{cc}
0 & {\overline {\widehat G}^{-1}} \\ 
{\widehat G}^{-1} & 0
\end{array}
\right].
\end{equation}
Let the notation $g^{{\overline a}b}$ stand for the kernel of the inverse operator ${\widehat G}^{-1}$ and let $g^{a{\overline b}}$ stand for its conjugate ${\overline g^{{\overline a}b}}$. Then
\begin{equation}
({\widehat G}_{R})^{ab}=({\widehat G}_{R})^{{\overline a}{\overline b}}=0,
\end{equation}
and
\begin{equation}
({\widehat G}_{R})^{{\overline a}b}=g^{{\overline a}b},
({\widehat G}_{R})^{a{\overline b}}={\overline g^{{\overline a}b}}.
\end{equation}
Accordingly, without danger of confusion we can denote the kernel of ${\widehat G}^{-1}_{R}$ by $g^{AB}$.

Having the Riemannian metric $G_{R}$ on $L_{2}$ we can define the compatible ({\it Riemannian}, or {\it Levi-Chevita}) connection $\Gamma$ by
\begin{equation}
\label{Levi}
2G_{R}(\Gamma(X,Y),Z)=dG_{R}X(Y,Z)+dG_{R}Y(Z,X)-dG_{R}Z(X,Y),
\end{equation}
for all vector fields $X,Y,Z$ in $H_{R}$. Here, for example, the term $dG_{R}X(Y,Z)$ denotes the derivative of the inner product $G_{R}(Y,Z)$ evaluated on the vector field $X$. 
In the given realization of the tangent bundle, for any $\varphi \in L_{2}$ the connection $\Gamma$ is an element of the space $L(H_{R},H_{R};H_{R})$. 
The latter notation means that $\Gamma$ is an $H_{R}$-valued $2$-form on $H_{R}\times H_{R}$.  
In our index notation the equation (\ref{Levi}) can be written as
\begin{equation}
2g_{AB}\Gamma^{B}_{CD}=\frac{\delta g_{AD}}{\delta \varphi^{C}}+\frac{\delta g_{CA}}{\delta \varphi^{D}}-\frac{\delta g_{CD}}{\delta \varphi^{A}}.
\end{equation}
Here for any $\varphi \in L_{2}$ the expression $g_{AB}\Gamma^{B}_{CD}$ is an element of $L(H_{R},H_{R},H_{R}; R)$, i.e., it is an $R$-valued $3$-form defined by
\begin{equation}
g_{AB}\Gamma^{B}_{CD}X^{C}Y^{D}Z^{A}=G_{R}(\Gamma(X,Y),Z)
\end{equation}
for all $X,Y,Z \in H_{R}$. Similarly, for any $\varphi \in L_{2}$, the functional derivative $\frac{\delta g_{AD}}{\delta \varphi^{C}}$ is an element of $L(H_{R},H_{R},H_{R}; R)$ defined by 
\begin{equation}
\frac{\delta g_{AD}}{\delta \varphi^{C}}X^{C}Y^{D}Z^{A}=dG_{R}X(Y,Z).
\end{equation}

For any $\varphi \in L_{2}$, by leaving vector $Z$ out, we can treat both sides of (\ref{Levi}) as elements of $H^{\ast}$.
Recall now that $G_{R}$ is a strong Riemannian metric. That is, for any $\varphi \in L_{2}$ the operator ${\widehat G}_{R}:H_{R} \longrightarrow H^{\ast}_{R}$ is an isomorphism, i.e., ${\widehat G}^{-1}_{R}$ exists. 
By applying ${\widehat G}^{-1}_{R}$ to both sides of (\ref{Levi}) without $Z$  we have in the index notation:
\begin{equation}
\label{Christoffel}
2\Gamma^{B}_{CD}=g^{BA}\left ( \frac{\delta g_{AD}}{\delta \varphi^{C}}+\frac{\delta g_{CA}}{\delta \varphi^{D}}-\frac{\delta g_{CD}}{\delta \varphi^{A}}\right ),
\end{equation}
where 
\begin{equation}
\Gamma^{B}_{CD}X^{C}Y^{D}\Omega_{B}=({\widehat G}^{-1}_{R}({\widehat G}_{R}\Gamma(X,Y)), \Omega).
\end{equation}
Formula (\ref{Christoffel}) defines the connection ``coefficients'' (Christoffel symbols) of the Levi-Chevita connection. From the matrix form of ${\widehat G}_{R}$ and ${\widehat G}^{-1}_{R}$ we can now easily obtain
\begin{equation}
\label{11}
\Gamma^{b}_{cd}={\overline \Gamma}^{{\overline b}}_{{\overline c}{\overline d}}=\frac{1}{2}g^{{\overline a}b}\left (\frac {\delta g_{d{\overline a}}}{\delta \varphi^{c}}+\frac {\delta g_{c{\overline a}}}{\delta \varphi^{d}}\right ),
\end{equation}
\begin{equation}
\label{22}
\Gamma^{b}_{c{\overline d}}={\overline \Gamma}^{{\overline b}}_{{\overline c}d}=\frac{1}{2}g^{{\overline a}b}\left (\frac {\delta g_{c{\overline a}}}{\delta {\overline \varphi} ^{d}}-\frac {\delta g_{c{\overline d}}}{\delta {\overline \varphi}^{a}}\right),
\end{equation}
\begin{equation}
\label{33}
\Gamma^{b}_{{\overline c}d}={\overline \Gamma}^{{\overline b}}_{c{\overline d}}=\frac{1}{2}g^{{\overline a}b}\left (\frac {\delta g_{d{\overline a}}}{\delta {\overline \varphi} ^{c}}-\frac {\delta g_{{\overline c}d}}{\delta {\overline \varphi}^{a}}\right),
\end{equation}
while the remaining components vanish. To compute the coefficients, let us write the metric (\ref{index_metric}) in the form
\begin{equation}
g_{a{\overline b}}=\frac{k_{a{\overline b}}}{\delta_{u{\overline v}}\varphi^{u} {\overline \varphi}^{v}},
\end{equation}
where $\delta_{u{\overline v}}\equiv \delta(u-v)$ is the $L_{2}$ metric in the index notation.
We then have for the derivatives:
\begin{equation}
\frac{\delta g_{a{\overline b}}}{\delta \varphi^{c}}=-\frac{k_{a{\overline b}}\delta_{c{\overline v}}{\overline \varphi}^{v}}{\left\|\varphi \right\|^{4}_{L_{2}}},
\end{equation}
and
\begin{equation}
\frac{\delta g_{a{\overline b}}}{\delta {\overline \varphi}^{c}}=-\frac{k_{a{\overline b}}\delta_{u{\overline c}}\varphi^{u}}{\left\|\varphi \right\|^{4}_{L_{2}}}.
\end{equation}
Using (\ref{11})-(\ref{33}) we can now find the non-vanishing connection coefficients
\begin{equation}
\label{111}
\Gamma^{b}_{cd}={\overline \Gamma}^{{\overline b}}_{{\overline c}{\overline d}}=-\frac{\left (\delta^{b}_{d}\delta_{c{\overline v}}+\delta^{b}_{c}\delta_{d{\overline v}}\right ){\overline \varphi}^{v}}{2\left\|\varphi \right\|^{2}_{L_{2}}},
\end{equation}
\begin{equation}
\label{222}
\Gamma^{b}_{c{\overline d}}={\overline \Gamma}^{{\overline b}}_{{\overline c}d}=-\frac{\left (\delta^{b}_{c}\delta_{u{\overline d}}-k^{{\overline a}b}k_{c{\overline d}}\delta_{u{\overline a}}\right )\varphi^{u}}{2\left\|\varphi \right\|^{2}_{L_{2}}},
\end{equation}
and
\begin{equation}
\label{333}
\Gamma^{b}_{{\overline c}d}={\overline \Gamma}^{{\overline b}}_{c{\overline d}}=-\frac{\left (\delta^{b}_{d}\delta_{u{\overline c}}-k^{{\overline a}b}k_{d{\overline c}}\delta_{u{\overline a}}\right )\varphi^{u}}{2\left\|\varphi \right\|^{2}_{L_{2}}}.
\end{equation}

Consider now the unit sphere $S^{L_{2}}: \left\|\varphi\right\|_{L_{2}}=1$ in the space $L_{2}$. Let ${\widehat A}$ be a (possibly unbounded) injective Hermitian operator defined on a set $D\left({\widehat A}\right)$ and with the image ${\widehat A}\left(D \right )\equiv R\left({\widehat A}\right)$. Here we assume for simplicity that $D\left({\widehat A}\right) \subset R\left({\widehat A}\right)$  and that both $D\left({\widehat A}\right)$ and $R\left({\widehat A}\right)$ are dense subsets of $L_{2}$. Let us define the inner product $(f,g)_{H}$ of any two elements $f,g$ in $R\left({\widehat A}\right)$ by the formula $\left(f,g\right)_{H} \equiv \left({\widehat A}^{-1}f, {\widehat A}^{-1}g\right)_{L_{2}}=\left (\left ({\widehat A}{\widehat A}^{\ast}\right)^{-1}f, g \right)$. By completing $R\left({\widehat A}\right)$ with respect to this inner product we obtain a Hilbert space $H$. Notice that ${\widehat A}$ is bounded in this norm and can be therefore extended to the entire space $L_{2}$ and becomes an isomorphism from $L_{2}$ onto $H$. We will denote such an extension by the same symbol ${\widehat A}$. Let ${\widehat K}=({\widehat A}{\widehat A}^{\ast})^{-1}$, ${\widehat K}: H \longrightarrow H^{\ast}$ be the metric operator on $H$. As before, we define the Riemannian metric on $L_{2\ast}$ by
\begin{equation}
\label{RiemR}
G_{R}(X,Y)=\frac{2Re({\widehat K}\xi,\eta)}{(\varphi,\varphi)_{L_{2}}},
\end{equation}
where $X=(\xi,{\overline \xi})$, $Y=(\eta,{\overline \eta})$.
Assume that the sphere $S^{L_{2}} \subset L_{2\ast}$ is furnished with the induced Riemannian metric. Consider now the vector field $A_{\varphi}=-i{\widehat A}\varphi$ associated with the operator ${\widehat A}$. The integral curves of this vector field are solutions of the equation $\frac{d \varphi_{t}}{dt}=-i{\widehat A}\varphi_{t}$. These solutions are given by $\varphi_{\tau}=e^{-i{\widehat A}\tau}\varphi_{0}$. Since $e^{-i{\widehat A}\tau}$ denotes a  one-parameter group of unitary operators, the integral curve $\varphi_{\tau}$ through a point $\varphi_{0} \in S^{L_{2}}$ stays on $S^{L_{2}}$ . In particular, the vector field $A_{\varphi}$ is tangent to the sphere. In other words, the operator $-i{\widehat A}$ maps points on the sphere into vectors tangent to the sphere.

We claim now that the curves $\varphi_{\tau}=e^{-i{\widehat A}\tau}\varphi_{0}$ are geodesics on the sphere in the induced metric. That is, they satisfy the equation
\begin{equation}
\label{geophi}
\frac{d^{2}\varphi_{\tau}}{d\tau^{2}}+\Gamma\left (\frac{d\varphi_{\tau}}{d\tau},\frac{d\varphi_{\tau}}{d\tau}\right)=0.
\end{equation}
In fact, using (\ref{111})-(\ref{333}) and collecting terms, we obtain
\begin{equation}
\label{geo1}
\Gamma^{b}_{CD}\frac{d\varphi^{C}_{\tau}}{d\tau}\frac{d\varphi^{D}_{\tau}}{d\tau}=\frac{\left({\widehat K}\frac{d\varphi_{\tau}}{d\tau},\frac{d\varphi_{\tau}}{d\tau}\right ){\widehat A}^{2}\varphi^{b}}{\left\| \varphi_{\tau}\right\|^{2}_{L_{2}}}.
\end{equation}
The expression for $\Gamma^{{\overline b}}_{CD}\frac{d\varphi^{C}_{\tau}}{d\tau}\frac{d\varphi^{D}_{\tau}}{d\tau}$ turns out to be the complex conjugate of (\ref{geo1}).
Now, the substitution of $\varphi_{\tau}=e^{i{\widehat A}\tau}\varphi_{0}$ and ${\widehat K}=\left ({\widehat A}{\widehat A}^{\ast}\right)^{-1}$ into the right hand side of (\ref{geo1}) yields ${\widehat A}^{2}\varphi$. At the same time, $\frac{d^{2}\varphi_{\tau}}{d\tau^{2}}=-{\widehat A}^{2}\varphi$ and therefore the equation (\ref{geophi}) is satisfied.
That is, the curves $\varphi_{\tau}=e^{-i{\widehat A}\tau}\varphi_{0}$ are geodesics in the metric (\ref{RiemR}) on $L_{2\ast}$. Since these curves also belong to the sphere $S^{L_{2}}$ and the metric on the sphere is induced by the embedding $S^{L_{2}} \longrightarrow L_{2\ast}$, we conclude that the curves $\varphi_{\tau}$ are geodesics on $S^{L_{2}}$.
 
Assume in particular that ${\widehat A}$ is the Hamiltonian ${\widehat h}$ of a closed quantum system. Then the above model demonstrates that, in the appropriate Riemannian metric on the unit sphere $S^{L_{2}}$, the Schr{\"o}dinger evolution of the system is a motion along a geodesic of $S^{L_{2}}$.

%%%%%%%%%%%%%%%%%%%%%
%%%%%%%%%%%%%%%%%%%%%

\end{document}